\documentclass{article}
\usepackage{hyperref}
\usepackage{fullpage}
\usepackage{footnote}
\usepackage{natbib}
\makesavenoteenv{algorithm}

\usepackage{amsmath} % text inside formulas ++
\usepackage{amsfonts} % enables mathbb
\usepackage{commath} % allows abs command
\usepackage{amsthm} % proofs
\usepackage{mathrsfs} %images
\usepackage{xcolor}
\usepackage{enumitem}
\usepackage{algorithm} % algorithms
\usepackage[noend]{algpseudocode}
\usepackage{xcolor}%comments
\usepackage[margin=0.5cm]{subcaption}  %figures side by side
\usepackage{mathtools} %:=
\usepackage{booktabs}
\usepackage{nicefrac}

\newtheorem{theorem}{Theorem}

{
\theoremstyle{definition}

\newtheorem{problem}{Problem}

\newtheorem*{remark*}{Remark}

}

\newcommand{\revset}{\mathcal{R}}
\newcommand{\papset}{\mathcal{W}}

\newcommand{\numrev}{m}
\newcommand{\numpap}{n}

\newcommand{\papload}{\lambda}
\newcommand{\revload}{\mu}

\newcommand{\revidx}{i}
\newcommand{\papidx}{j}
\newcommand{\conflicts}{C}
\newcommand{\authorship}{A}
\newcommand{\assignment}{M}
\newcommand{\ranking}{\pi}

\newcommand{\aggrule}{\Lambda}

\newcommand{\hypothesis}{H}

\newcommand{\permutation}{p}

\newcommand{\specrev}{\revidx^{\ast}}
\newcommand{\identity}{I}

\newcommand{\impranking}{\ranking^{\ast}}

\newcommand{\commentedout}[1]{}
\newcommand{\efsize}{\Delta}

\newcommand{\uniform}{\mathcal{U}}
\newcommand{\expectation}{\mathbb{E}}

\newcommand{\val}{v}

\newcommand{\numparticipants}{N}
\newcommand{\noise}{\sigma}
\newcommand{\distance}{\texttt{Distance}}

\newcommand{\reverse}{\texttt{Reverse}}
\newcommand{\heuristic}{\texttt{See-Saw}}
\newcommand{\response}{\texttt{2x-Distance}}
\newcommand{\first}{\texttt{Better-to-Bottom}}
\newcommand{\second}{\texttt{Worse-to-Bottom}}
\newcommand{\numpart}{55}
\newcommand{\valset}{\Phi}
\newcommand{\stat}{\tau}
\newcommand{\level}{\alpha}
\newcommand{\conflictset}{\mathcal{P}}
\newcommand{\tmpconflicts}{C'}
\newcommand{\tmpstat}{\varphi}
\newcommand{\tmpranking}{\ranking'}
\newcommand{\randranking}{\tilde{\ranking}}
\newcommand{\tmpauthorship}{\authorship'}
\newcommand{\Perm}{\Pi}

\title{Catch Me if I Can:\\ Detecting Strategic Behaviour in Peer Assessment}
\author{\\
  Ivan Stelmakh, Nihar B. Shah and Aarti Singh\\~\\
  School of Computer Science \\ 
  Carnegie Mellon University\\
  \texttt{\{stiv,nihars,aarti\}@cs.cmu.edu} 
}
\date{}

\begin{document}

\maketitle

\begin{abstract}
We consider the issue of strategic behaviour in various peer-assessment tasks, including peer grading of exams or homeworks and peer review in hiring or promotions. When a peer-assessment task is competitive (e.g., when students are graded on a curve), agents may be incentivized to misreport evaluations in order to improve their own final standing. Our focus is on designing methods  for detection of such manipulations. Specifically, we consider a setting in which agents evaluate a subset of their peers and output rankings  that are later aggregated to form a final ordering. In this paper, we investigate a statistical framework for this problem and design a principled test for detecting strategic behaviour. We prove that our test has strong false alarm guarantees and evaluate its detection ability in practical settings. For this, we design and execute an experiment that elicits strategic behaviour from subjects and release a dataset of patterns of strategic behaviour that may be of independent interest. We then use the collected data to conduct a series of real and semi-synthetic evaluations that demonstrate a strong detection power of our test. 

\end{abstract}
%%%%%%%%%%%%%%%%%%%%%%%%%%%%%%%%%%%%
\section{Introduction}

Ranking a set of items submitted by a group of people (or ranking the people themselves) is a ubiquitous task that is faced in many applications, including education, hiring, employee evaluation and promotion, and academic peer review. Many of these applications have a large number of submissions, which makes obtaining an evaluation of each item by a set of independent experts prohibitively expensive or slow. Peer-assessment techniques offer an appealing alternative: instead of relying on independent judges, they distribute the evaluation task across the fellow applicants and then aggregate the received reviews into the final ranking of items. This paradigm has become popular for employee evaluation~\citep{edwards96threesixzero} and grading students' homeworks~\citep{topping98peerassessment}, and is now expanding to more novel applications of massive open online courses~\citep{kulkarni13peerassessment, piech13moocs} and hiring at scale~\citep{kotturi19HirePeer}.    

The downside of such methods, however, is that reviewers are incentivized to evaluate their counterparts strategically to ensure a better outcome of their own item~\citep{huang2019discovery, balietti2016peer, hassidim18admission}. Deviations from the truthful behaviour decrease the overall quality of the resulted ranking and undermine fairness of the process. This issue has led to a long line of work~\citep{alon09sumofus, aziz16peers, kurokawa15impartial, kahng18ranking, xu2019strategyproofArxiv} on designing ``impartial'' aggregation rules that can eliminate the impact of the ranking returned by a reviewer on the final position of their item.

While impartial methods remove the benefits of manipulations, such robustness may come at the cost of some accuracy loss when reviewers do not engage in strategic behaviour. This loss is caused by less efficient data usage~\citep{kahng18ranking, xu2019strategyproofArxiv} and reduction of efforts put by reviewers~\citep{kotturi19HirePeer}. Implementation of such methods also introduces some additional logistical burden on the system designers; as a result, in many critical applications (e.g., conference peer review) the non-impartial mechanisms are employed. An important barrier that prevents stakeholders from making an informed choice to implement an impartial mechanism is a lack of tools to detect strategic behaviour. Indeed, to evaluate the trade off between the loss of accuracy due to manipulations and the loss of accuracy due to impartiality, one needs to be able to evaluate the extent of strategic behaviour in the system. With this motivation, in this work we \emph{focus on detecting strategic manipulations in peer-assessment processes.}

Specifically, in this work we consider a setting where each reviewer is asked to evaluate a subset of works submitted by their counterparts. In a carefully designed randomized study of strategic behaviour when evaluations take the form of \emph{ratings}, \citet{balietti2016peer} were able to detect manipulations by comparing the distribution of scores given by target reviewers to some truthful reference. However, other works~\citep{huang2019discovery, barroga2014safeguarding} suggest that in more practical settings reviewers may strategically decrease some scores and increase others in attempt to mask their manipulations or intentionally promote weaker submissions, thereby keeping the distribution of output scores unchanged and making the distribution-based detection inapplicable. Inspired by this observation, we aim to design tools to detect manipulations when the distributions of scores output by reviewers are fixed, that is, we assume that evaluations are collected in the form of \emph{rankings}. Ranking-based evaluation is used in practice~\citep{hazelrigg2013DearCL} and has some theoretical properties that make it appealing for peer grading~\citep{shah13acf, caragiannis14ranking} which provides additional motivation for our work. 

\noindent \textbf{Contributions} In this work we present two sets of results.

\begin{itemize}[itemsep=2pt, leftmargin=*, topsep=0pt]
    \item \textbf{Theoretical.} First, we propose a non-parametric test for detection of strategic manipulations in peer-assessment setup with rankings. Second, we prove that our test has a reliable control over the false alarm probability (probability of claiming existence of the effect when there is none).  Conceptually, we avoid difficulties associated to dealing with rankings as covariates by carefully accounting for the fact that each reviewer is ``connected'' to their submission(s);  therefore, the manipulation they employ is naturally not an arbitrary deviation from the truthful strategy, but instead the deviation that potentially improves the outcome of their works. 
    
    \item \textbf{Empirical.} On the empirical front, we first design and conduct an experiment that incentivizes strategic behaviour of participants. This experiment yields a novel dataset of patterns of strategic behaviour that we make publicly available and that can be useful for other researchers (the dataset is available on the first author's website). Second, we use the experimental data to evaluate the detection power of our test on answers of real participants and in a series of semi-synthetic simulations. These evaluations demonstrate that our testing procedure has a non-trivial detection power, while not making strong modelling assumptions on the manipulations employed by strategic agents. 

\end{itemize}

The remainder of this paper is organized as follows. We give a brief overview of related literature in Section~\ref{sec:lit}. In Section~\ref{sec:preliminaries} we formally present the problem setting and demonstrate an important difference between the ranking and rating setups. Next, in Section~\ref{section:test} we design a novel approach to testing for strategic behaviour and prove the false alarm guarantees for our test. Section~\ref{sec:experiment} is dedicated to the discussion of the experiment that we designed and executed to collect a dataset of patterns of strategic behaviour. In Section~\ref{section:evaluation} we use this dataset to evaluate the detection ability of our test. Finally, we conclude the paper with a discussion in Section~\ref{section:discussion}.

%%%%%%%%%%%%%%%%%%%%%%%%%%%%%%%%%%%%%%%%%
\section{Related Literature}
\label{sec:lit}

Our work falls at the intersection of crowdsourcing, statistical testing, and a recent line of computer science research on the peer-assessment process. We now give an overview of relevant literature from these areas.

\smallskip

\noindent \textbf{Crowdsourcing work on manipulations.} Despite motivation for this work comes from studies of~\citet{balietti2016peer} and~\citet{huang2019discovery}, an important difference between rankings and ratings that we highlight in Section~\ref{section:comparison} makes the models considered in these works inapplicable to our setup. Several other papers~\citep{thurner2011peer, cabota2013competition, paolucci2014mechanism} specialize on the problem of strategic behaviour in peer review and perform simulations to explore its detrimental impact on the quality of published works. These works are orthogonal to the present paper because they do not aim to detect the manipulations. 

Another relevant paper~\citep{perez18distopt} considers a problem of strategic behaviour in context of the relationships between electric vehicle aggregators in the electricity market. In that setup, each agent is supposed to solve a part of a certain distributed optimization problem and self-interested agents may be incentivized to misreport their solutions. \citet{perez18distopt} offer a heuristic procedure to identify manipulating agents, but the proposed method relies on the specific nature of the optimization problem and does not directly extend to our setup. 

Finally, a long line of work~\citep{akoglu13fraud, kaghazgaran18review, jindal08spam} aims at detecting fraud in online consumer reviews. In contrast to our setting, these works try to identify malicious reviews without having a direct and known connection between the reviewers and the items being reviewed that is present in our setup. Instead, these works often rely on additional information (e.g., product-specific features) which is irrelevant or unavailable in our problem.

\smallskip

\noindent \textbf{Statistical testing.} In this paper, we formulate the test for strategic behaviour as a test for independence of rankings returned by reviewers from their own items. Classical statistical works~\citep{lehmann2005testing} for independence testing are not directly applicable to this problem due to the absence of low-dimensional representations of items. To avoid dealing with unstructured items, one could alternatively formulate the problem as a two-sample test and obtain a control sample of rankings from non-strategic reviewers. This approach, however, has two limitations. First, past work suggests that the test and control rankings may have different distributions even under the absence of manipulations due to misalignment of incentives~\citep{kotturi19HirePeer}. Second, existing works~\citep{mania18kernels, gretton12kernel, jiao18permutations, rastogi20pairwise} on two-sample testing with rankings ignore the authorship information that is crucial in our case as we show in the sequel (Section~\ref{section:comparison}).

\smallskip

\noindent \textbf{Research on peer assessment.} This paper also falls in the line of several recent works in computer science on the peer-evaluation process that includes both empirical~\citep{tomkins17wsdm, sajjadi16peergrading, kotturi19HirePeer} and theoretical~\citep{wang18calibration, stelmakh2018pr4a, noothigattu2018choosing,fiez2020super} studies. Particularly relevant works are recent papers~\citep{tomkins17wsdm, stelmakh2019test} that consider the problem of detecting biases (e.g., gender bias) in single-blind peer review. Biases studied therein manifest in reviewers being harsher to some subset of submissions (e.g., papers authored by females), making the methods designed in these works inapplicable to the problem we study. Indeed, in our case there does not exist a fixed subset of works that reviewers need to put at the bottom of their rankings to improve the outcome of their own submissions. However, these works share a conceptual approach of detecting the effect on the aggregate level of all agents rather than in each agent individually.

As discussed earlier, research on peer review also aims to prevent or mitigate strategic behavior, where reviewers may manipulate their reviews to help their own papers~\citep{alon09sumofus, aziz16peers, kurokawa15impartial, kahng18ranking, xu2019strategyproofArxiv} or manipulate reviews to maliciously influence the outcomes of other papers~\citep{jecmen2020mitigating}.

%%%%%%%%%%%%%%%%%%%%%%%%%%%%%%%%%%%%%%%%%

\section{Problem Formulation}
\label{sec:preliminaries}

In this section we present our formulation of the manipulation-testing problem.

%%%%%%%%%%%%%%%%%%%%%%%%%%%%%%%%%%%%%%%%%

\subsection{Preliminaries}
\label{sec:setting}

In this paper, we operate in the peer-assessment setup in which reviewers first conduct some work (e.g., homework assignments) and then judge the performance of each other. We consider a setting where reviewers are asked to provide a total ranking of the set of works they are assigned to review.

We let $\revset = \{1, 2, \ldots, {\numrev}\}$ and $\papset = \{1, 2, \ldots, {\numpap}\}$ denote the set of reviewers and works submitted for review, respectively. We let matrix $\conflicts \in \{0, 1\}^{\numrev \times \numpap}$ represent conflicts of interests between reviewers and submissions, that is, $(\revidx, \papidx)^{\text{th}}$ entry of $\conflicts$  equals 1 if reviewer $\revidx$ is in conflict with work $\papidx$ and 0 otherwise. Matrix $\conflicts$ captures all kinds of conflicts of interest, including authorship, affiliation and others, and many of them can be irrelevant from the manipulation standpoint (e.g., affiliation may put a reviewer at conflict with dozens of submissions they are not even aware of). We use $\authorship \in \{0, 1\}^{\numrev \times \numpap}$ to denote a subset of ``relevant'' conflicts --- those that reviewers may be incentivized to manipulate for --- identified by stakeholders. For the ease of presentation, we assume that $\authorship$ represents the authorship conflicts, as reviewers are naturally interested in improving the final standing of their own works, but in general it can capture any subset of conflicts. For each reviewer $\revidx \in \revset$, non-zero entries of the corresponding row of matrix $\authorship$ indicate submissions that are (co-)authored by reviewer $\revidx$. We let $\conflicts(\revidx)$ and $\authorship(\revidx) \subseteq \conflicts(\revidx)$ denote possibly empty sets of works conflicted with and authored by reviewer $\revidx$, respectively. 

Each work submitted for review is assigned to $\papload$ non-conflicting reviewers subject to a constraint that each reviewer gets assigned $\revload$ works. For brevity, we assume that parameters $\numpap, \numrev, \revload, \papload$ are such that $\numpap \papload = \numrev \revload$ so we can assign exactly $\revload$ works to each reviewer. The assignment is represented by a binary matrix $\assignment \in \{ 0, 1\}^{\numrev \times \numpap}$ whose $(\revidx, \papidx)^{\text{th}}$ entry equals 1 if reviewer $\revidx$ is assigned to work $\papidx$ and 0 otherwise. We call an assignment valid if it respects the (submission, reviewer)-loads and does not assign a reviewer to a conflicting work. Given a valid assignment $\assignment$ of works $\papset$ to reviewers $\revset$, for each $\revidx \in \revset$, we use $\assignment({\revidx})$ to denote a set of works assigned to reviewer $\revidx$.  $\Perm\left[\assignment(\revidx)\right]$ denotes a set of all $|\assignment(\revidx)|!$ total rankings of these works and reviewer $\revidx$ returns a ranking $\ranking_{\revidx} \in \Perm\left[\assignment(\revidx)\right]$. The rankings from all reviewers are aggregated to obtain a final ordering $\aggrule(\ranking_1, \ranking_{2}, \ldots, \ranking_{\numrev})$ that matches each work $\papidx \in \papset$ to its position $\aggrule_{\papidx}(\ranking_1, \ranking_{2}, \ldots, \ranking_{\numrev})$, using some aggregation rule $\aggrule$ known to all reviewers. The grades or other rewards are then distributed according to the final ordering $\aggrule(\ranking_1, \ranking_{2}, \ldots, \ranking_{\numrev})$ with authors of higher-ranked works receiving better grades or rewards.

In this setting, reviewers may be incentivized to behave strategically because the ranking they output may impact the outcome of \emph{their own} works. The focus of this work is on designing tools to detect strategic behaviour of reviewers when a non-impartial aggregation rule $\aggrule$ (e.g., a rule that theoretically allows reviewers to impact the final standing of their own submissions) is employed. 

%%%%%%%%%%%%%%%%%%%%%%%%%%%%%%%%%%%%%%%%%

\subsection{Motivating Example}
\label{section:comparison}

To set the stage, we start from highlighting an important difference between rankings and ratings in the peer-assessment setup. To this end, let us consider the experiment conducted by~\citet{balietti2016peer} in which reviewers are asked to give a score to each work assigned to them for review and the final ranking is computed based on the mean score received by each submission. It is not hard to see that in their setting, the dominant strategy for each rational reviewer who wants to maximize the positions of their own works in the final ranking is to give the lowest possible score to all submissions assigned to them. Observe that this strategy is fixed, that is, it does not depend on the quality of reviewer's work --- irrespective of position of their work in the underlying ordering, each reviewer benefits from assigning the lowest score to all submissions they review. Similarly, \citet{huang2019discovery} in their work also operate with ratings and consider a fixed model of manipulations in which strategic agents increase the scores of low-quality submissions and decrease the scores of high-quality submissions, irrespective of the quality of reviewers' works.

In contrast, when reviewers are asked to output \emph{rankings} of submissions, the situation is different and reviewers can no longer rely on fixed strategies to gain the most for their own submission. To highlight this difference, let us consider a toy example of the problem with 5 reviewers and 5 submissions ($\numrev = \numpap = 5$), authorship and conflict matrix given by an identity matrix ($\conflicts = \authorship = \identity$), and three works (reviewers) assigned to each reviewer (work), that is, $\papload = \revload = 3$. In this example, we additionally assume that: (i) assignment of reviewers to works is selected uniformly at random from the set of all valid assignments, (ii) aggregation rule $\aggrule$ is the Borda count, that is, the positional scoring rule with weights equal to positions in the ranking,\footnote{We use the variant without tie-breaking --- tied submissions share the same position in the final ordering.}  (iii) reviewers are able to reconstruct the ground-truth ranking of submissions assigned to them
without noise, and (iv) all but one reviewers are truthful. 

\begin{figure}[ht]
    \centering
    \includegraphics[width=0.5\textwidth]{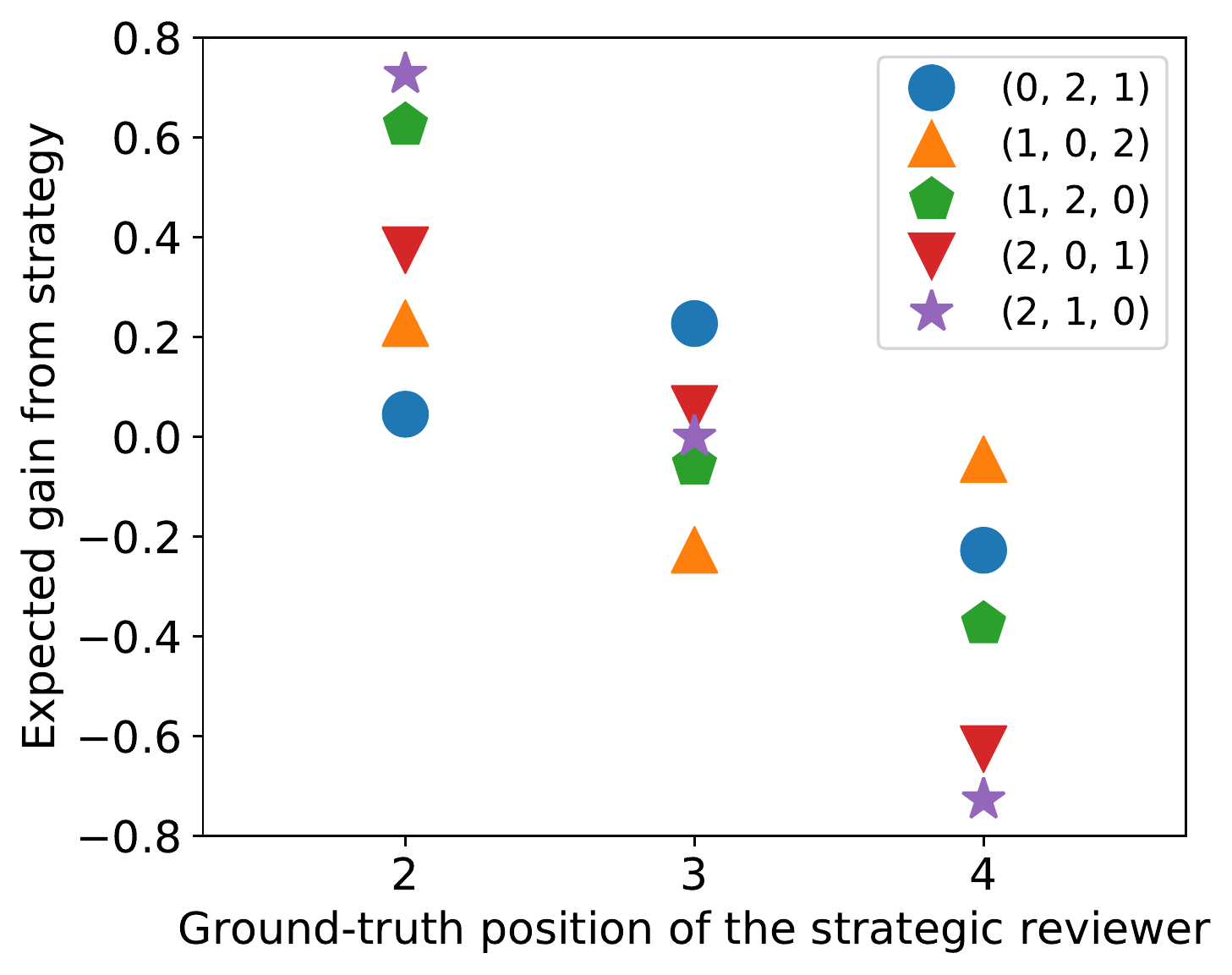}
    \caption{Comparison of fixed deterministic strategies available to a single strategic reviewer depending on position of their work in the true underlying ranking. A positive value of the expected gain indicates that the manipulation strategy in expectation delivers better position in the final ordering than the truthful strategy.}
    \label{fig:onestrategin}
\end{figure}

Under this simple formulation, we qualitatively analyze the strategies available to the strategic reviewer, say reviewer $\specrev$. Specifically, following the rating setup, we consider the fixed deterministic strategies that do not depend on the work created by reviewer $\specrev$. Such strategies are limited to permutations of the ground-truth ranking of submissions in $\assignment({\specrev})$. Figure~\ref{fig:onestrategin} represents an expected gain of each strategy (measured in positions in the aggregated ordering) as compared to the truthful strategy for positions 2--4 of the work authored by reviewer $\specrev$ in the ground-truth ranking. To make this figure, for each target position of the strategic reviewer  $\specrev$ in the underlying total ordering, we first compute their expected position in the final ranking (expectation is taken over the randomness in the assignment) if they use the truthful strategy. We then compute the same expectations for each manipulating strategy and plot the differences. The main observation is that there does not exist a fixed strategy that dominates the truthful strategy for every possible position of the reviewer's work. Therefore, in setup with rankings strategic reviewers need to consider how their own works compare to the works they rank in order to improve the outcome of their submissions.

%%%%%%%%%%%%%%%%%%%%%%%%%%%%%%%%%%%%%

\subsection{Problem Setting}
\label{sec:formulation}

With the motivation given in Section~\ref{section:comparison}, we are ready to present the formal hypothesis-testing problem we consider in this work. When deciding on how to rank the works, the information available to reviewers is the content of the works they review and the content of their own works. Observe that while a truthful reviewer does not take into account their own submissions when ranking works of others, the aforementioned intuition suggests that the ranking output by a strategic agent should depend on their own works. Our formulation of the test for manipulations as an independence test captures this motivation.

\begin{problem}[Testing for strategic behaviour]
\label{problem:independence}
 Given a non-impartial aggregation rule $\aggrule$, assignment of works to reviewers $\assignment$, rankings returned by reviewers $\left\{\ranking_\revidx, \revidx \in \revset \right\}$, conflict matrix $\conflicts$, authorship matrix $\authorship$ and set of works submitted for review $\papset$, the goal is to test the following hypotheses:
 
 \smallskip
 
\noindent \textbf{Null} $(\hypothesis_0)$
\begin{align*}
   \forall \revidx \in \revset \ \text{s.t.} \ \authorship(\revidx) \neq \emptyset \quad \ranking_{\revidx} \perp \authorship(\revidx).
\end{align*} 

 \smallskip

\noindent \textbf{Alternative}  $(\hypothesis_1)$ 
\begin{align*}
    \exists \revidx \in \revset \ \text{s.t.} \ \authorship(\revidx) \neq \emptyset \quad \ranking_{\revidx} \not\perp \authorship(\revidx).
\end{align*}
\end{problem}
In words, under the null hypothesis reviewers who have their submissions under review do not take into account their own works when evaluating works of others and hence are not engaged in manipulations that can improve the outcome of their own submissions. In contrast, under the alternative hypothesis some reviewers choose the ranking depending on how their own works compare to works they rank, suggesting that they are engaged in manipulations.

\smallskip 

\noindent \textbf{Assumptions} Our formulation of the testing problem makes two assumptions about the data-generation process to ensure that association between works authored by reviewer $\revidx$ and ranking $\ranking_{\revidx}$ may be caused only by strategic manipulations and not by some intermediate mediator variables. 

\begin{enumerate}[itemsep=0pt, topsep=0pt, leftmargin=*, label=(A\arabic*)]
    \item \label{assumption:assignment} \textbf{Random assignment.} We assume that the assignment of works to reviewers is selected uniformly at random from the set of all assignments that respect the conflict matrix $\conflicts$. This assumption ensures that the works authored by a reviewer do not impact the set of works assigned to them for review. The assumption of random assignment holds in many applications, including peer grading~\citep{freeman10accurate, kulkarni13peerassessment} and NSF review of proposals~\citep{hazelrigg2013DearCL}.

    \item \label{assumption:noise} \textbf{Independence of ranking noise.} We assume that under the null hypothesis of absence of strategic behaviour, the reviewer identity is independent of the works they author, that is, the noise in reviewers' evaluations (e.g., the noise due to subjectivity of the truthful opinion) is not correlated with their submissions. This assumption is satisfied by various popular models for generation of rankings, including Plackett-Luce model~\citep{luce59model, plackett75model} and more general location family of random utility models~\citep{soufiani12rums}.
\end{enumerate}

In the sequel, we show that Assumption A1 can be relaxed to allow for assignments of any fixed topology (Appendix~\ref{appendix:relaxation}) and that our test can control the false alarm probability under some practical violations of Assumption A2 (Appendix~\ref{appendix:moreeval}). More generally, we note that one needs to carefully analyze the assumptions in the specific application and carefully interpret the results of the test, keeping in mind that its interpretation depends heavily on whether the assumptions are satisfied or not.

%%%%%%%%%%%%%%%%%%%%%%%%%%%%%%%%%%%%%%%%%

\section{Testing Procedure} 
\label{section:test}

In this section, we introduce our testing procedure. Before we delve into details, we highlight the main intuition that determines our approach to the testing problem. Observe that when a reviewer engages in strategic behaviour, they tweak their ranking to ensure that \emph{their own} works experience better outcome when all rankings are aggregated by the rule $\aggrule$. Hence, when \emph{successful} strategic behaviour is present, we may expect to see that the ranking returned by a reviewer influences position of \emph{their own} works under aggregation rule $\aggrule$ in a more positive way than other works not reviewed by this reviewer. Therefore, the test we present in this paper attempts to identify whether rankings returned by reviewers have a more positive impact on the final standing of their own works than what would happen by chance.

For any reviewer $\revidx \in \revset$, let $\uniform_{\revidx}$ be a uniform distribution over rankings $\Perm\left[\assignment(\revidx)\right]$ of works assigned to them for review. With this notation, we formally present our test as Test~\ref{test:withcontrasts} below. Among other arguments, our test accepts the optional set of rankings $\{ \impranking_{\revidx}, \revidx \in \revset \}$, where for each $\revidx \in \revset$,  $\impranking_{\revidx}$ is a ranking of works $\assignment(\revidx)$ assigned to reviewer $\revidx$, but is constructed by an impartial agent (e.g., an outsider reviewer who does not have their own works in submission). For the ease of exposition, let us first discuss the test in the case when the optional set of rankings is \emph{not} provided (i.e., the test has no supervision) and then we will make a case for usefulness of this set.

In Step~\ref{test1:stat}, the test statistic is computed as follows: for each reviewer $\revidx \in \revset$ and for each work ${\papidx} \in \authorship(\revidx)$ authored by this reviewer, we compute the impact of the ranking returned by the reviewer on the final standing of this work. To this end, we compare the position actually taken by the work (first term in the inner difference in Equation~\ref{eqn:statcontrast}) to the expected position it would take if the reviewer would sample the ranking of works $\assignment(\revidx)$ uniformly at random (second term in the inner difference in Equation~\ref{eqn:statcontrast}). To get the motivation behind this choice of the test statistic, note that if a reviewer $\revidx$ is truthful then the ranking they return may be either better or worse for \emph{their own} submissions than a random ranking, depending on how their submissions compare to works they review. In contrast, a strategic reviewer may choose the ranking that delivers a better final standing for their submissions, biasing the test statistic to the negative side.

Having defined the test statistic, we now understand its behaviour under the null hypothesis to quantify when its value is too large to be observed under the absence of manipulations for a given significance level $\level$. To this end, we note that for a given assignment matrix $\assignment$, there are many pairs of conflict and authorship matrices $(\tmpconflicts, \tmpauthorship)$ that (i) are equal to the actual matrices $\conflicts$ and $\authorship$ up to permutations of rows and columns and (ii) do not violate the assignment $\assignment$, that is, do not declare a conflict between any pair of reviewer $\revidx$ and submission $\papidx$ such that submission ${\papidx}$ is assigned to reviewer $\revidx$ in $\assignment$. Next, note that under the null hypothesis of absence of manipulations, the behaviour of reviewers would not change if matrix $\authorship$ was substituted by another matrix $\tmpauthorship$, that is, a ranking returned by any reviewer $\revidx$ would not change if that reviewer was an author of works $\tmpauthorship(\revidx)$ instead of $\authorship(\revidx)$. Given that the structure of the alternative matrices $\tmpconflicts$ and $\tmpauthorship$ is the same as that of the actual matrices $\conflicts$ and $\authorship$, under the null hypothesis of absence of manipulations we expect the actual test statistic to have a similar value as compared to that under $\tmpconflicts$ and $\tmpauthorship$.

{
\setcounter{algorithm}{0}
\floatname{algorithm}{Test}
\begin{algorithm*}[h]
   \caption{Test for strategic behaviour}
   \label{test:withcontrasts}
   {\bfseries Input:} Reviewers' rankings $\{\ranking_\revidx, \revidx \in \revset\}$ \\
   \hphantom{{\bfseries Input:}} Assignment $\assignment$ of works to reviewers \\ 
   \hphantom{{\bfseries Input:}} Conflict and authorship matrices $(\conflicts, \authorship)$  \\ 
   \hphantom{{\bfseries Input:}} Significance level $\level$, aggregation rule $\aggrule$\\
   {\bfseries Optional Argument:} Impartial rankings $\{\impranking_{\revidx},  \revidx \in \revset\}$

    \begin{enumerate}[leftmargin=*, topsep=4pt]
    
        \item \label{test1:stat} Compute the test statistic $\stat$ as
        \begin{align}
        \label{eqn:statcontrast}
            \stat = \sum\limits_{\revidx \in \revset} \sum\limits_{\papidx \in \authorship(\revidx)} \Big( \aggrule_{{\papidx}}(\tmpranking_{1}, \tmpranking_{2}, \ldots, \ranking_i, \ldots, \tmpranking_{\numrev}) - \expectation_{\randranking \sim \uniform_{\revidx}} \left[ \aggrule_{{\papidx}}(\tmpranking_{1}, \tmpranking_{2}, \ldots, \randranking, \ldots, \tmpranking_{\numrev}) \right] \Big),
        \end{align}
        where $\tmpranking_{\revidx}, \revidx \in \revset,$ equals $\impranking_{\revidx}$ if the optional argument is provided and equals $\ranking_{\revidx}$ otherwise.
        
        \item \label{test1:distr} Compute a multiset $\conflictset(\assignment)$ as follows. For each pair ($\permutation_{\numrev}, \permutation_{\numpap}$) of permutations of $\numrev$ and $\numpap$ items, respectively, apply permutation $\permutation_{\numrev}$ to rows of matrices $\conflicts$ and $\authorship$ and permutation $\permutation_{\numpap}$ to columns of matrices $\conflicts$ and $\authorship$. Include the obtained matrix $\tmpauthorship$ to $\conflictset(\assignment)$ if it holds that for each $\revidx \in \revset$:
        \begin{align*}
            \tmpauthorship(\revidx) \subseteq \tmpconflicts(\revidx) \subset \papset \backslash \assignment({\revidx}).
        \end{align*}
        
        \item \label{test1:sample} For each matrix $\tmpauthorship \in \conflictset(\assignment)$ define $\tmpstat(\tmpauthorship)$ to be the value of the test statistic~\eqref{eqn:statcontrast} if we substitute $\authorship$ with $\tmpauthorship$, that is, $\tmpstat(\tmpauthorship)$ is the value of the test statistic if the authorship relationship was represented by $\tmpauthorship$ instead of $\authorship$. Let 
        \begin{align}
        \label{eqn:multiset}
            \valset = \left\{ \tmpstat(\tmpauthorship), \tmpauthorship \in \conflictset(\assignment) \right\}
        \end{align} 
        denote  the multiset that contains all these values.
        
        \item \label{test1:decision} Reject the null if $\tau$ is strictly smaller than the $(\left \lfloor {\level \abs{\valset}} \right \rfloor + 1)^{\text{th}}$ order statistic of $\Phi$. 
    \end{enumerate}
\end{algorithm*}}

The aforementioned idea drives Steps~\ref{test1:distr}-\ref{test1:decision} of the test. In Step~\ref{test1:distr} we construct the set of all pairs of conflict and authorship matrices of the fixed structure that do not violate the assignment $\assignment$. We then compute the value of the test statistic for each of these authorship matrices in Step~\ref{test1:sample} and finally reject the null hypothesis in Step~\ref{test1:decision} if the actual value of the test statistic $\stat$ appears to be too extreme against values computed in Step~\ref{test1:sample} for the given significance level $\level$.

If additional information in the form of impartial rankings is available (i.e., the test has a supervision), then our test can detect manipulations better. The idea of supervision is based on the following intuition. In order to manipulate successfully, strategic reviewers need to have some information about the behaviour of others. In absence of such information, it is natural (and this idea is supported by data we obtain in the experiment in Section~\ref{sec:experiment}) to choose a manipulation targeted against the truthful reviewers, assuming that a non-trivial fraction of agents behave honestly. The optional impartial rankings allow the test to use this intuition: for each reviewer $\revidx \in \revset$ the test measures the impact of reviewer's ranking on their submissions as if this reviewer was the only manipulating agent, by complementing the ranking $\ranking_{\revidx}$ with impartial rankings $\{\impranking_{1},  \ldots, \impranking_{\revidx-1}, \impranking_{\revidx + 1} \ldots, \impranking_{\numrev} \}$. As we show in Section~\ref{section:evaluation}, availability of supervision can significantly aid the detection power of the test.

The following theorem combines the above intuitions and ensures a reliable control over the false alarm probability for our test (the proof is given in Appendix~\ref{appendix:proof}).

\begin{theorem}
\label{thm:type1contrasts}
    Suppose that Assumptions A1 and A2 specified in Section~\ref{sec:formulation} hold. Then, under the null hypothesis of absence of manipulations, for any significance level $\level \in (0, 1)$ and for any aggregation rule $\aggrule$, Test~\ref{test:withcontrasts} (both with and without supervision) is guaranteed to reject the null with probability at most $\level$. Therefore, Test~\ref{test:withcontrasts} controls the false alarm probability at the level $\alpha$.
\end{theorem}

\begin{remark*}
    1. In Section~\ref{section:evaluation} we complement the statement of the theorem by demonstrating that our test has a non-trivial detection power. 
    
    2. In practice, the multiset $\conflictset(\assignment)$ may take $\mathcal{O}\left(\numrev!\numpap!\right)$ time to construct which is prohibitively expensive even for small values of $\numrev$ and $\numpap$. The theorem holds if instead of using the full multiset $\conflictset(\assignment)$, when defining $\valset$, we only sample some $k$ authorship matrices uniformly at random from the multiset $\conflictset(\assignment)$. The value of $k$ should be chosen large enough to ensure that $(\left \lfloor {\level |\valset|} \right \rfloor + 1)$ is greater than  $1$. The sampling can be performed by generating random permutations using the shuffling algorithm of~\citet{fisher65permutation} and rejecting samples that lead to matrices $\tmpauthorship \notin \conflictset(\assignment)$.
    
    3. The impartial set of rankings $\{\impranking_i, i \in \revset \}$ need not necessarily be constructed by a separate set of $\numrev$ reviewers. For example, if one has access to the (noisy) ground-truth (for example, to the  ranking of homework assignments constructed by an instructor), then for each $\revidx \in \revset$ the ranking $\impranking_{i}$ can be a ranking of $\assignment({\revidx})$ that agrees with the ground-truth. 
\end{remark*}

\noindent \textbf{Effect size} In addition to controlling for the false alarm probability, our test offers a measure of the effect size defined as $\efsize = \stat \cdot \left[\sum_{\revidx \in \revset} |\authorship(\revidx)| \right]^{-1}$. Each term in the test statistic $\stat$ defined in~\eqref{eqn:statcontrast} captures the impact of the ranking returned by a reviewer on the final standing of the corresponding submission and the the mean impact is a natural measure of the effect size. Negative values of the effect size demonstrate that reviewers in average benefit from the rankings they return as compared to rankings sampled uniformly at random.

%%%%%%%%%%%%%%%%%%%%%%%%%%%%%%%
\section{Experiment to Elicit Strategic Behaviour}
\label{sec:experiment}

In this section we describe the experiment we designed and executed to collect a dataset of patterns of strategic behaviour that we will use in Section~\ref{section:evaluation} to empirically evaluate the detection power of our test. The experiment was offered to attendees of a graduate-level AI course at Carnegie Mellon University and $\numparticipants = \numpart$ students completed the experimental procedure described in Section~\ref{sec:design}. Exploratory analysis of the collected data is given in Section~\ref{section:eda} and the dataset is available on the first author's website.

%%%%%%%%%%%%%%%%%%%%%%%%%%%%%%%%%%%
\subsection{Design of Experiment}
\label{sec:design}

The goal of our experiment is to understand what strategies people use when manipulating their rankings of others. A real peer grading setup (i.e., homework grading) possesses an ethical barrier against cheating and hence many subjects of the hypothetical experiment would behave truthfully, reducing the efficiency of the process. To overcome this issue, we use gamification and organize the experiment as follows (game interface can be found in supplementary materials on the first author's website).

We design a game for $\numrev = 20$ players and $\numpap = 20$ hypothetical submissions. First, a one-to-one authorship relationship $\authorship$ is sampled uniformly at random from the set of permutations of 20 items and each player becomes an ``author'' of one of the submissions. Each submission is associated to a unique value $\val \in \{1, 2, \ldots, 20 \}$ and this value is privately communicated to the respective player; therefore, players are associated to values and in the sequel we do not distinguish between a player's value and their ``submission''. We then communicate values of some $\revload = 4$ other contestants to each player subject to the constraint that a value of each player becomes known to $\papload = 4$ counterparts. To do so, we sample an assignment $\assignment$ from the set of assignments respecting the conflict matrix $\conflicts = \authorship$ uniformly at random. Note that players do not get to see the full assignment and only observe the values of players assigned to them. The rest of the game replicates the peer grading setup: participants are asked to rank their peers (the truthful strategy is to rank by values in decreasing order) and the rankings are aggregated using the Borda count aggregation rule (tied submissions share the position in the final ordering).

For the experiment, we create 5 rounds of the game, sampling a separate authorship matrix $\authorship_{k}$ and assignment $\assignment_{k}$ for each round $k \in \{1, 2, \ldots, 5\}$. Each of the $\numparticipants = \numpart$ subjects then participates in all 5 rounds, impersonating one (the same for all rounds) of the 20 game players.\footnote{We sample a separate authorship matrix for each round so participants get different values between rounds.} Importantly, subjects are instructed that their goal is to \emph{manipulate their ranking to improve their final standing}. Additionally, we inform participants that in the first 4 rounds of the game their competitors are truthful bots who always rank players by their values. In the last round, participants are informed that they play against other subjects who also engage in manipulations.

To help participants better understand the rules of the game and properties of the aggregation mechanism, after each of the first four rounds, participants are given feedback on whether their strategy improves their position in the aggregated ordering. Note that the position of the player in the final ordering depends on the complex interplay between (i) the strategy they employ, (ii) the strategy employed by others, and (iii) the configuration of the assignment. In the first four rounds of the game, participants have the information about (ii), but do not get to see the third component. To make feedback independent of (iii), we average it out by computing the mean position of the player over the randomness in the part of the assignment unobserved by the player and give positive feedback if their strategy is in expectation better than the ranking sampled uniformly at random. Finally, after the second round of the game, we give a hint that additionally explains some details of the game mechanics.

The data we collect in the first four rounds of the game allows us to understand what strategies people use when they manipulate in the setup when (most) other reviewers are truthful. In the last round, we remove the information about the behaviour of others and collect data about manipulations in the wild (i.e., when players do not know other players' strategies). Manual inspection of the collected data reveals that 53 participants attempted manipulations in each round and the remaining 2 subjects manipulated in all but one round each, hence, we conclude that the data is collected under the alternative hypothesis of the presence of manipulations.

%%%%%%%%%%%%%%%%%%%%%%%%%%%%%%%%%%%
\subsection{Exploratory Data Analysis}
\label{section:eda}

We now continue to the exploratory analysis of collected data and begin from analyzing the manipulating strategies employed by participants. In addition to rankings, in each round we asked participants to describe their reasoning in a textual form and we manually analyze these descriptions to identify the strategies people use. While these textual descriptions sometimes do not allow to unequivocally understand the general strategy of the player due to ambiguity, we are able to identify 6 broad clusters of strategies employed by participants. We underscore that each of these clusters may comprise several strategies that are similar in spirit but may slightly disagree in some situations. We now introduce these clusters by describing the most popular representative strategy that will be used in the subsequent analysis.

\begin{itemize}[topsep=5pt, leftmargin=15pt, itemsep=0.05em]
    \item \reverse{} This naive strategy prescribes to return the reversed ground truth ordering of players under comparison. Note that in contrast to other strategies we explain below, the ranking returned by a player who use this strategy is independent of their own value. 
    
    \item \distance{} The idea behind this family of strategies is to identify the direct competitors and put them at the bottom of the ranking, while out of reach players and those with considerably smaller values are put at the top. The most popular incarnation of this strategy is to rank the other players in order of decreasing distance from \emph{the player's} value: the furthest player gets the first place and the closest gets the last place. 
    
    \item \heuristic{} This strategy follows \reverse{} if the value assigned to a player is in top 50\% of all values (i.e., greater than $10$) and follows the truthful strategy otherwise. None of the participants directly reported this strategy in the experiment, but we include it in the analysis as this strategy agrees with behaviour of several players.
    
    \item \first{} This strategy is another simplification of the \distance{} strategy. Let $\val^{\ast}$ be the player's value. Then this strategy prescribes to put submissions with values smaller than $\val^{\ast}$ at the top (in order of increasing values) and submissions with values larger than $\val^{\ast}$ at the bottom (in order of decreasing values) of the ranking. For example, if the player's value is 10 and they are asked to rank other players whose values are $(16, 12, 7, 2)$, then this strategy would return $\ranking = 2 \succ 7 \succ 16 \succ 12$.
    
    \item \second{} Submissions with values lower than $\val^{\ast}$ are placed at the bottom (in order of decreasing values) and submissions with values larger than $\val^{\ast}$ are placed at the top (in order of increasing values) of the ranking. In the earlier example with $\val^{\ast} = 10$ and values $(16, 12, 7, 2)$ to be ranked, this strategy would return $\ranking = 12 \succ 16 \succ 7 \succ 2$.
    
    \item \response{} This strategy was reported only in round 5, that is, when participants were competing against each other, and is targeted to respond to the \distance{} strategy. This strategy suggests redefining all values (including the value of the player) by the following rule:
    \begin{align*}
        \val' = \min\{20 - \val, \val - 1 \},
    \end{align*}
    and apply the \distance{} strategy over the updated values.
\end{itemize}

\begin{figure}
    \centering
    \includegraphics[width=9cm]{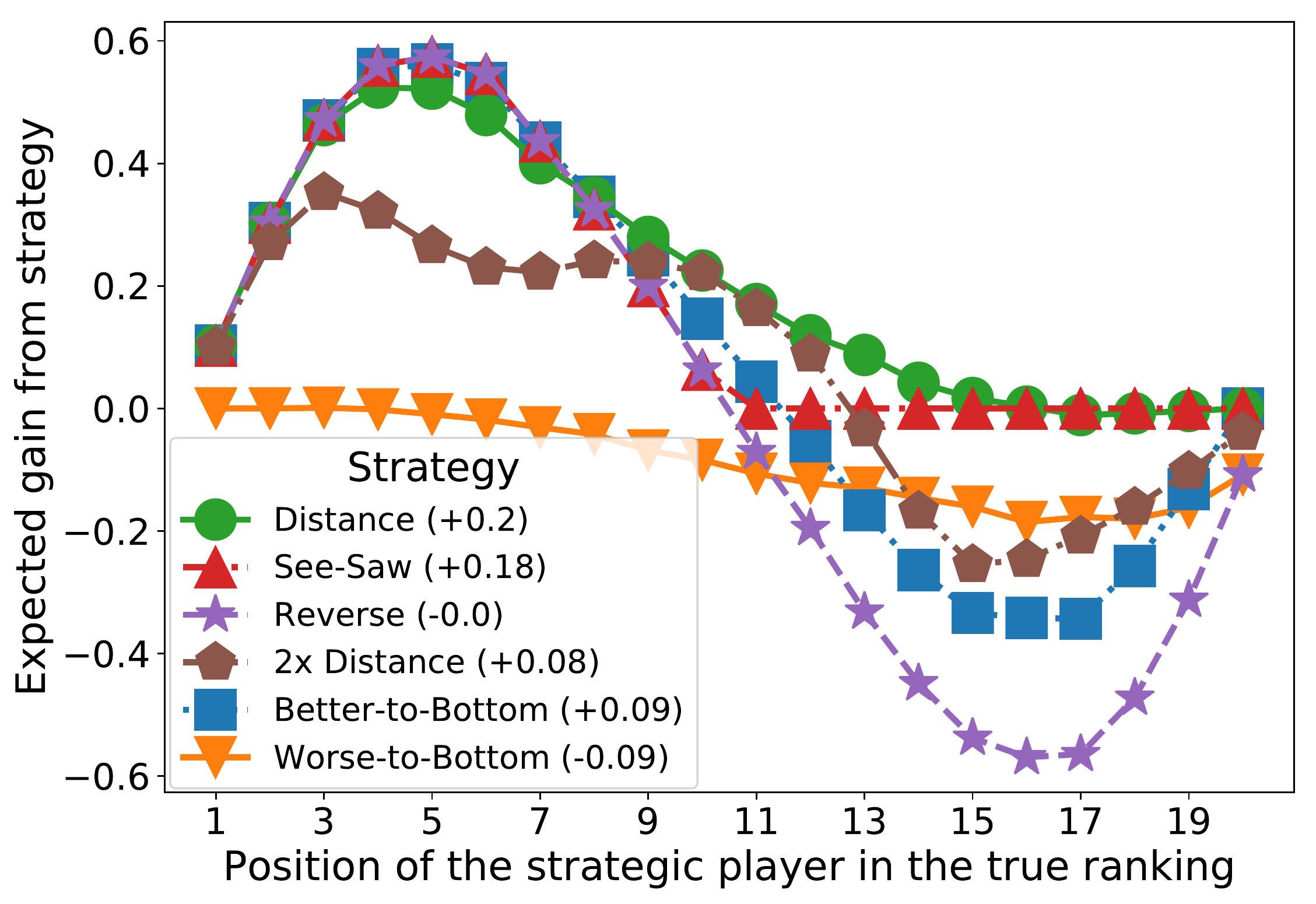}
    \caption{Expected gain from manipulation strategy when all but one player are truthful as a function of position of the strategic player in in the ground-truth ranking. The expectation is taken over randomness of the assignment procedure and values in brackets in the legend correspond to the mean gain over all positions. Borda count aggregation rule is used. A positive value of the expected gain indicates that the manipulation strategy in expectation delivers better position in the final ordering than the truthful strategy. Error bars are too small to be visible.}
    \label{fig:compare_strategies}
\end{figure}

Figure~\ref{fig:compare_strategies} juxtaposes the identified strategies by comparing them to the truthful one in case when all but one player are truthful. For each position of the strategic reviewer $\specrev$ in the ground-truth total ordering, we compute the expected gain (measured in positions in the aggregated ordering) from using each of the 6 strategies. To this end, we first compute the expected position (expectation is taken over randomness in the assignment) of reviewer $\specrev$ if they use the truthful strategy. We then compute the same expectations for each of the 6 manipulation strategies and plot the differences as a function of the position of the strategic player in the true underlying ranking.

We make several observations from Figure~\ref{fig:compare_strategies}. First, strategies \distance{} and \heuristic{} benefit the manipulating player irrespective of her/his position in the underlying ranking. In contrast, \first{} and \response{} can both help and hurt the player depending on the position in the ground-truth ordering and different effects average out to the positive total gain. The \reverse{} strategy delivers zero gain in expectation over positions, being not better nor worse than the truthful strategy. Finally, the \second{} strategy is uniformly dominated by the truthful strategy, implying that the strategic player can only hurt their expected position by relying on this strategy.

To conclude the preliminary analysis of collected data, for each of the 5 rounds we manually allocate each player to one of the aforementioned manipulation strategies based on the ranking and textual description they provided. As we mentioned above, this information is sometimes not sufficient to unequivocally identify the strategy. To overcome this ambiguity, we use fractional allocation in case several strategies match the response and leave some players unclassified in hard cases (for example, when textual response contradicts the actual ranking). Note that players who employed the truthful strategy are also included in the unclassified category as the goal of the categorization is to understand the behaviour of strategic players.

Table~\ref{table:strategies} displays the resulting allocation of players to strategies informed by the data collected in the experiment. First, in  round 1 of the game half of strategic players employed the \reverse{} strategy which is not better than the truthful strategy and hence does not lead to a successful manipulation. Second, as the game proceeds and players understand the mechanics of the game better, they converge to the \distance{} strategy which in expectation delivers a positive gain irrespective of the position of the player in the underlying ranking. Third, note that most of the players continued with the \distance{} strategy even in Round 5, despite in this round they were no longer playing against truthful bots. However, a non-trivial fraction of students managed to predict this behaviour and employed the \response{} strategy to counteract the \distance{} strategy. Finally, many players were clueless about what strategy to employ in round 5, contributing to the increased number of unclassified participants.

\begin{table}[htbp]
\vskip 0.15in
\begin{center}
\begin{small}
\begin{sc}
\begin{tabular}{lccccr}
\toprule
               &  Round 1 & Round 2 & Round 3 & Round 4 & Round 5     \\
\midrule
\reverse{}       & $.50$     & $.33$    & $.05$   & $.03$ & $.06$      \\
\distance{}      & $.37$    & $.53$    & $.93$   & $.96$ & $.78$      \\
\heuristic{}     & $.09$    & $.08$    & $.02$   & $.01$ & --         \\
\first{}         & $.02$    & $.04$    & --      & --    & --         \\
\second{}        & $.02$    & $.02$    & --      & --    & --         \\
\response{}      & --       & --       & --      & --    & $.16$      \\
\midrule
{Unclassified}  & 5        & 7        & 4       & 4     & 18 \\
\bottomrule       
\end{tabular}
\end{sc}
\end{small}
\end{center}
\caption{Manually encoded characterization of strategies used by manipulating participants. In the first 4 rounds of the game participants played against truthful bots and in the last round they played against each other.}
\label{table:strategies}
\end{table}

%%%%%%%%%%%%%%%%%%%%%%%%%%%%%%%%%%%
\section{Evaluation of the Test}
\label{section:evaluation}

We now investigate the detection power of our test (Test~\ref{test:withcontrasts}). We begin from analysis of real data collected in the previous section and execute the following procedure. For each of the 1,000 iterations, we uniformly at random subset 20 out of the \numpart{} participants such that together they impersonate all 20 game players. We then apply our test (with and without supervision) to rankings output by these participants in each of the 5 rounds, setting significance level at $\level = 0.05$ and sampling $k = 100$ authorship matrices in Step~\ref{test1:sample} of the test. The impartial rankings for testing with supervision comprise ground truth rankings.

\begin{table*}[ht]
\begin{center}
\begin{small}
\begin{sc}
\begin{tabular}{lccccc}
\toprule
               &  Round 1 & Round 2 & Round 3 & Round 4 & Round 5         \\
\midrule
With supervision & $0.61$ & $0.57$ & $0.87$ & $1.00$ & $0.09$  \\
Without supervision & $0.17$ & $0.02$ & $0.16$ & $0.01$ & $0.08$   \\
\bottomrule       
\end{tabular}
\end{sc}
\end{small}
\end{center}
\caption{Detection rates of our test.} 
\label{table:detection}
\end{table*}

After performing all iterations, for each round we compute the mean detection rate and represent these values in Table~\ref{table:detection}. The results suggest that our test provided with the impartial set of rankings has a strong detection power, reliably detecting manipulations in the first 4 rounds. On the other hand, performance of our test without supervision is modest. The reason behind the difference in performance is that our test aims at detecting \emph{successful} manipulations (i.e., those that improve the outcome of a player). In the first 4 rounds of the game, subjects were playing against truthful competitors and hence the test provided with the additional set of impartial rankings (which is targeted at detecting responses to the truthful strategy) has a good performance. However, the test without supervision is not able to detect such manipulations, because it evaluates success using rankings of other participants who also engage in manipulations and the response to the truthful strategy is not necessarily successful in this case. As for the round 5, we will show in a moment that poor performance of our test appears to be due to random chance (i.e., the choice of the assignment which is hard for detection) and not due to any systematic issue.

\begin{figure*}[ht]
\centering
\begin{subfigure}[t]{.48\textwidth}
  \centering  
  \includegraphics[height=5cm]{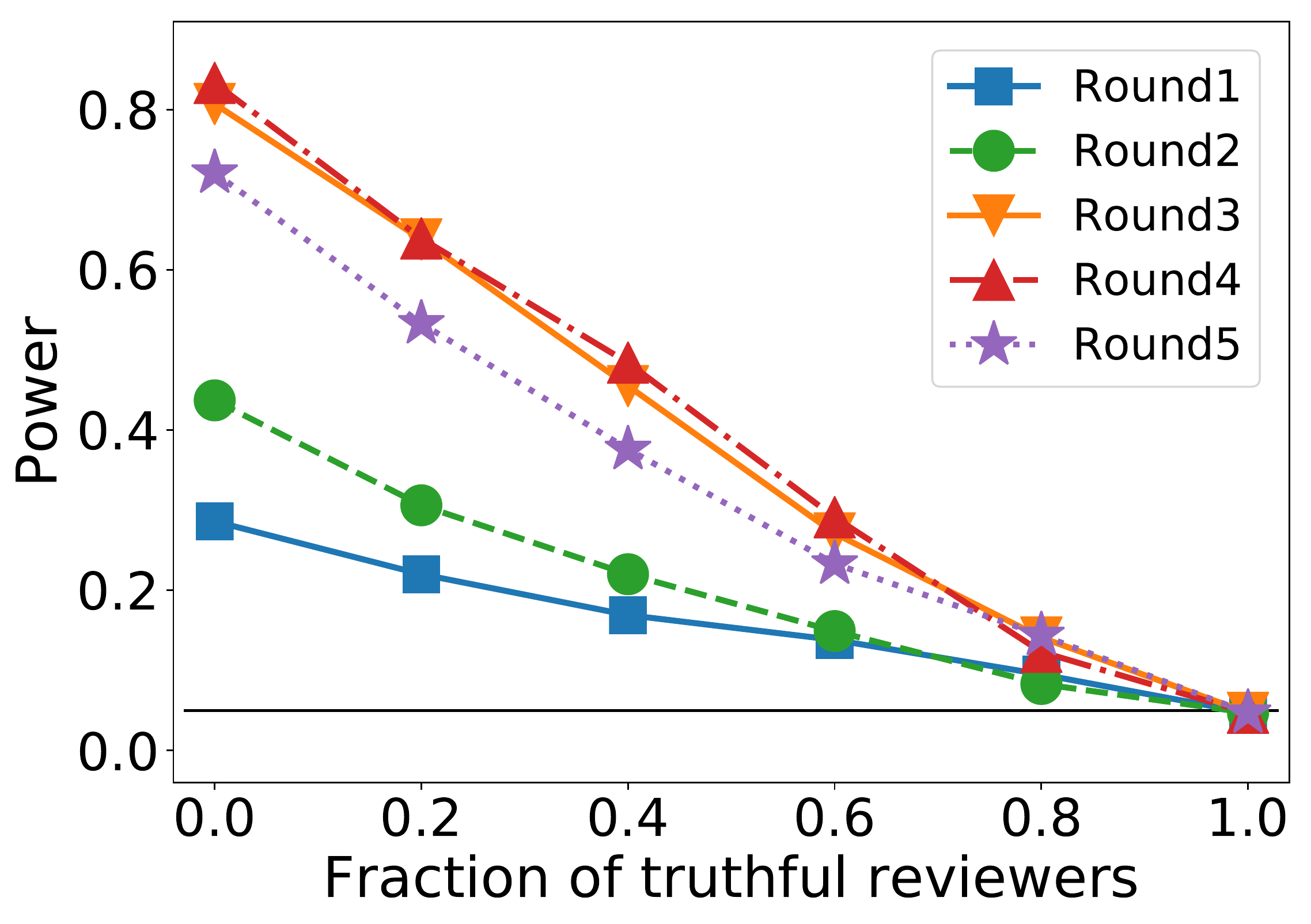}
  \caption{With supervision}
  \label{fig:power_synthetic:t1}
\end{subfigure}%
\begin{subfigure}[t]{.48\textwidth}
  \centering
  \includegraphics[height=5cm]{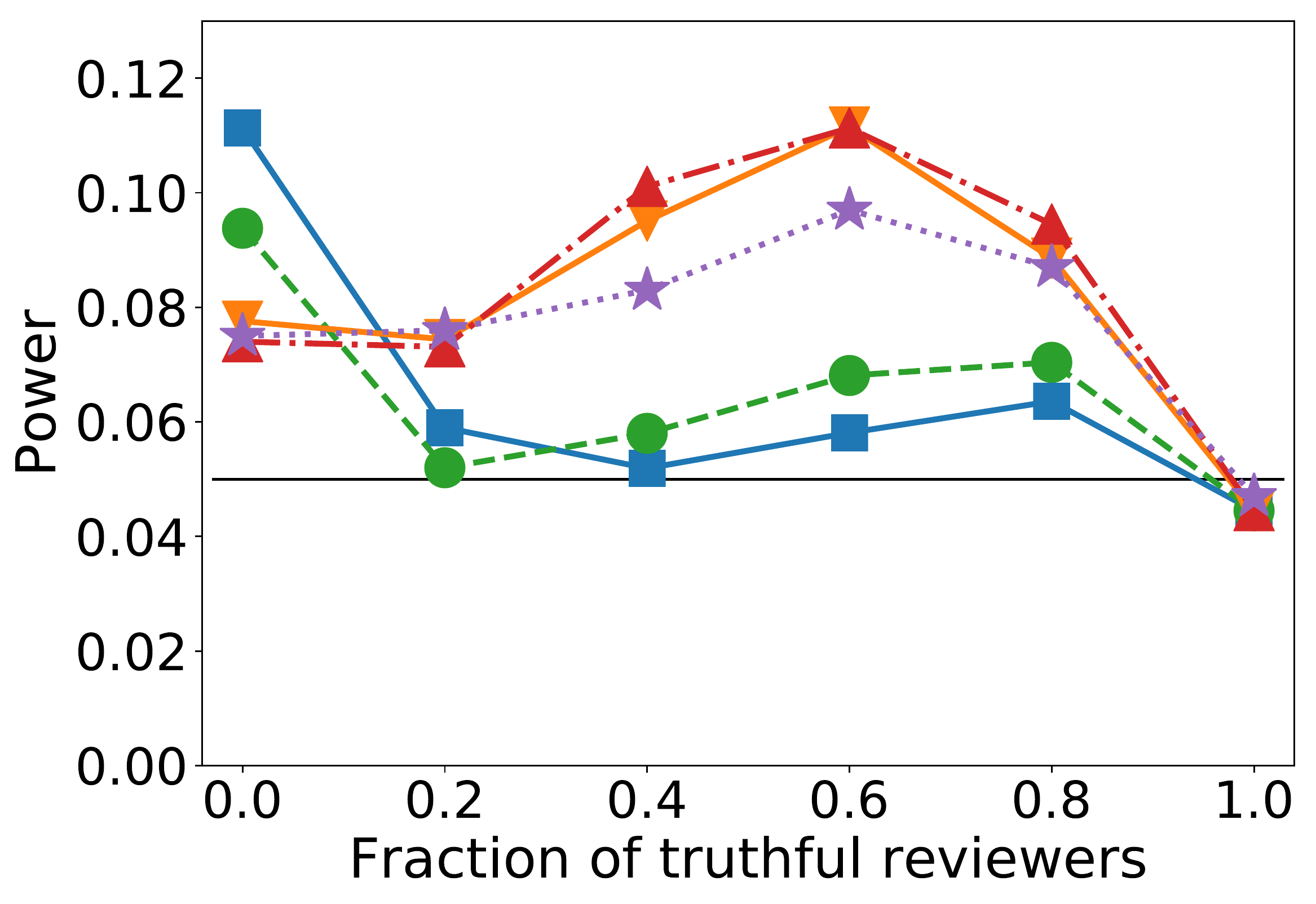}
  \caption{Without supervision}
  \label{fig:power_synthetic:t2}
\end{subfigure}
\caption{Expected power of our test for different allocations of strategic agents to strategies and different fractions of truthful agents. The black horizontal line is a baseline power achieved by a test that rejects the null with probability $\level{=}0.05$ irrespective of the data. Error bars are too small to show.}
\label{fig:power_synthetic}
\end{figure*}

Note that performance of our test depends not only on the strategies employed by players, but also on the assignment $\assignment$ realized in a particular round. Some realizations of random assignment make successful manipulations (and their detection) easier while under other realizations most of the players cannot improve their position even if they use the best strategy (and therefore our test cannot detect manipulations). To remove the impact of the specific assignments we used in the experiment, we now proceed to semi-synthetic trials. Specifically, we use the manual allocation of participants to manipulation strategies represented in Table~\ref{table:strategies} and create artificial agents who follow these strategies, replicating proportions learned from the real data. We then repeat our experiment with $\numrev = 20$ artificial agents, simulating 1,000 assignments for each round of the game and computing the expectation of the power of our test over randomness of the assignment. Additionally, we enhance the set of synthetic agents with truthful agents and study how the detection power of our test changes with the fraction of truthful agents.  Figure~\ref{fig:power_synthetic} displays the expected power of our test for proportions of strategies used by strategic agents informed by each round of the real game and for various fractions of truthful players. Note that when all players are truthful (rightmost points of both plots), the data is generated under the null hypothesis of absence of strategic behaviour, and the plots empirically verify the guarantee of Theorem~\ref{thm:type1contrasts} that our test indeed caps the false alarm rate at $\level = 0.05$.

Figure~\ref{fig:power_synthetic:t1} shows that our test provided with optional rankings has a non-trivial power in every round, including the last round in which participants were playing against each other. Note that as game proceeds and participants understand the rules better (and find ways to manipulate efficiently), the power of the test increases. A surprising success of the test with supervision in round 5 is explained by the combination of two factors: (i) the majority of participants resorted to the response to the truthful strategy even in round 5 and (ii) a strategy that constitutes a response to the response to the truthful strategy is still a good response to the truthful strategy. Hence, our test provided with impartial rankings can detect manipulations even in case when participants play against each other. 

Figure~\ref{fig:power_synthetic:t2} shows that the test without supervision has considerably lower (but still non-trivial) power. We note, however, that the main feature of the test without supervision is that it can be readily applied to purely observational data and the power can be accumulated over multiple datasets (e.g., it can be applied to multiple iterations of a university course). An interesting feature of the test without supervision is the non-monotonicity of power with respect to the fraction of truthful reviewers, caused by a complex interplay between the fraction of truthful agents and the strategies employed by manipulating agents that determines success of manipulations.

Overall, the evaluations we conducted in this section suggests that the test we propose in this work has a non-trivial detection power, especially when an additional set of impartial rankings is available. In Appendix~\ref{appendix:moreeval} we provide additional empirical evaluations of the test, including the case when reviewers are noisy and Assumption A2 (formulated in Section~\ref{sec:formulation}) is violated. 

%%%%%%%%%%%%%%%%%%%%%%%%%%%%%%%%

\section{Discussion} 
\label{section:discussion}

In this work, we design a test for detection of strategic behaviour in the peer-assessment setup with rankings. We prove that it has a reliable control over the false alarm probability and demonstrate its non-trivial detection power on data we collected in a novel experiment. Thus, our work offers a tool for system designers to measure the presence of strategic behavior in the peer-assessment system (peer-grading of homeworks and exams, evaluation of grant proposals, and hiring at scale) and informs the trade off between the loss of accuracy due to manipulations and the loss of accuracy due to restrictions put by impartial aggregation mechanisms. Therefore, organizers can employ our test to make an informed decision on whether they need to switch to the impartial mechanism or not. 

An important feature of our test is that it aims at detecting the manipulation on the aggregate level of all agents. As a result, our test does not allow for personal accusations and hence does not increase any pressure on individual agents. As a note of caution, we caveat, however, that selective application of our test (as well as of \emph{any} statistical test) to a specific sub-population of agents may lead to discriminatory statements; to avoid this, experimenters need to follow pre-specified experimental routines and consider ethical issues when applying our test.

Our approach is conceptually different from the past literature which considers ratings~\citep{balietti2016peer, huang2019discovery} as it does not assume any specific parametric model of manipulations and instead aims at detecting any \emph{successful} manipulation of rankings, thereby giving flexibility of non-parametric tests. This flexibility, however, does not extend to the case when agents try to manipulate but do it \emph{unsuccessfully} (see Appendix~\ref{appendix:moreeval} for demonstration). Therefore, an interesting problem for future work is to design a test that possesses flexibility of our approach but is also able to detect any (and not only successful) manipulations.

\section*{Acknowledgments}
This work was supported in part by NSF CAREER award 1942124 and in part by NSF CIF 1763734.

\bibliographystyle{apalike}
\bibliography{bibtex.bib}

\newpage

\appendix

\noindent \textbf{\LARGE{Appendix}}

\bigskip

\noindent We provide supplementary materials and additional discussion.  Appendix~\ref{appendix:moreeval} is dedicated to additional evaluations of our test (Test~\ref{test:withcontrasts}). We show how to slightly relax Assumption A1 of random assignment in Appendix~\ref{appendix:relaxation} and prove Theorem~\ref{thm:type1contrasts} in Appendix~\ref{appendix:proof}.

%%%%%%%%%%%%%%%%%%%%%%%%%%%%%%%%%%%
\section{Additional Evaluations of the Test}
\label{appendix:moreeval}

We now provide additional evaluations of our test and conduct simulations in the following settings:
\begin{itemize}[itemsep=0pt, leftmargin=*]
    \item \textbf{Detecting pure strategies.} First, we evaluate the detection power of the test against each of the strategies we learned from the experimental data (described in Section~\ref{section:eda}). 
    
    \item \textbf{Noisy supervision.} Next, we evaluate robustness of our test to the noise in the optional impartial rankings $\{\impranking_{\revidx}, \revidx \in \revset \}$.
    
    \item \textbf{Noise in reviewers' evaluations.} We also study the detection power of our test when reviewers perceive quality scores of submissions assigned to them with noise.
    
    \item \textbf{Violation of Assumption A2.} We then analyze the behavior of our test when Assumption A2 formulated in Section~\ref{sec:formulation} is violated. To this end, we use a model that connects the quality of reviewer's submission to the level of noise in their evaluations suggested by empirical research on peer grading and compute the false alarm rate of our test under this model.
    
    \item \textbf{Runtime of the test.} In this paper, we perform simulations in the small sample size setting ($\numpap = \numrev = 20$) to be able to run thousands of iterations and average out the impact of the specific assignment on the performance of our test. In practice, organizers will need to run the test only once (or several times if the test is applied over multiple datasets) and we provide runtimes of our naive implementation of the test for larger values of the problem parameters.
\end{itemize}

%%%%%%%%%%%%%%%%%%%%%%%%%%%%%%%
\subsection{Detecting Pure Strategies}
\label{appendix:moreeval:pure}

To compute the power against specific strategies we identified in Section~\ref{section:eda}, we follow the same approach we used to evaluate the expected detection power of our test in each round of the game (Figure~\ref{fig:power_synthetic}). Specifically, for each fraction of truthful agents and for each of the strategies we learned from data, we compute the detection power of the test with and without supervision over 1,000 assignments sampled uniformly at random from the set of all assignments valid for parameters:

\begin{align}
\label{eqn:simparam}
    \authorship = \conflicts = \identity, \ \numpap = \numrev = 20, \ \papload = \revload = 4.
\end{align}

\begin{figure}[ht]
\centering
\begin{subfigure}[t]{.48\textwidth}
  \centering  
  \includegraphics[height=4.5cm]{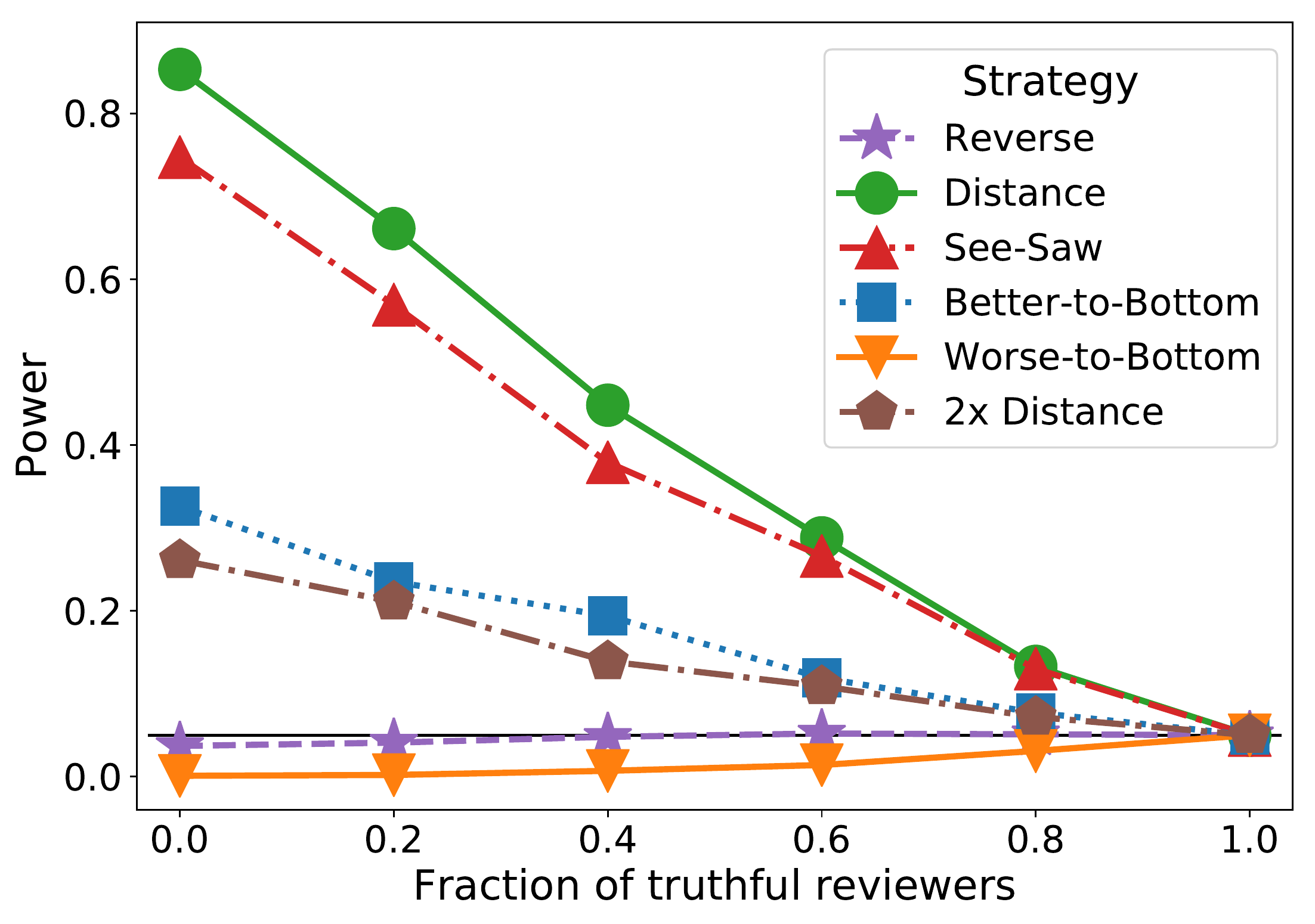}
  \caption{With supervision}
  \label{fig:power_synthetic_single:t1}
\end{subfigure}%
\begin{subfigure}[t]{.48\textwidth}
  \centering
  \includegraphics[height=4.5cm]{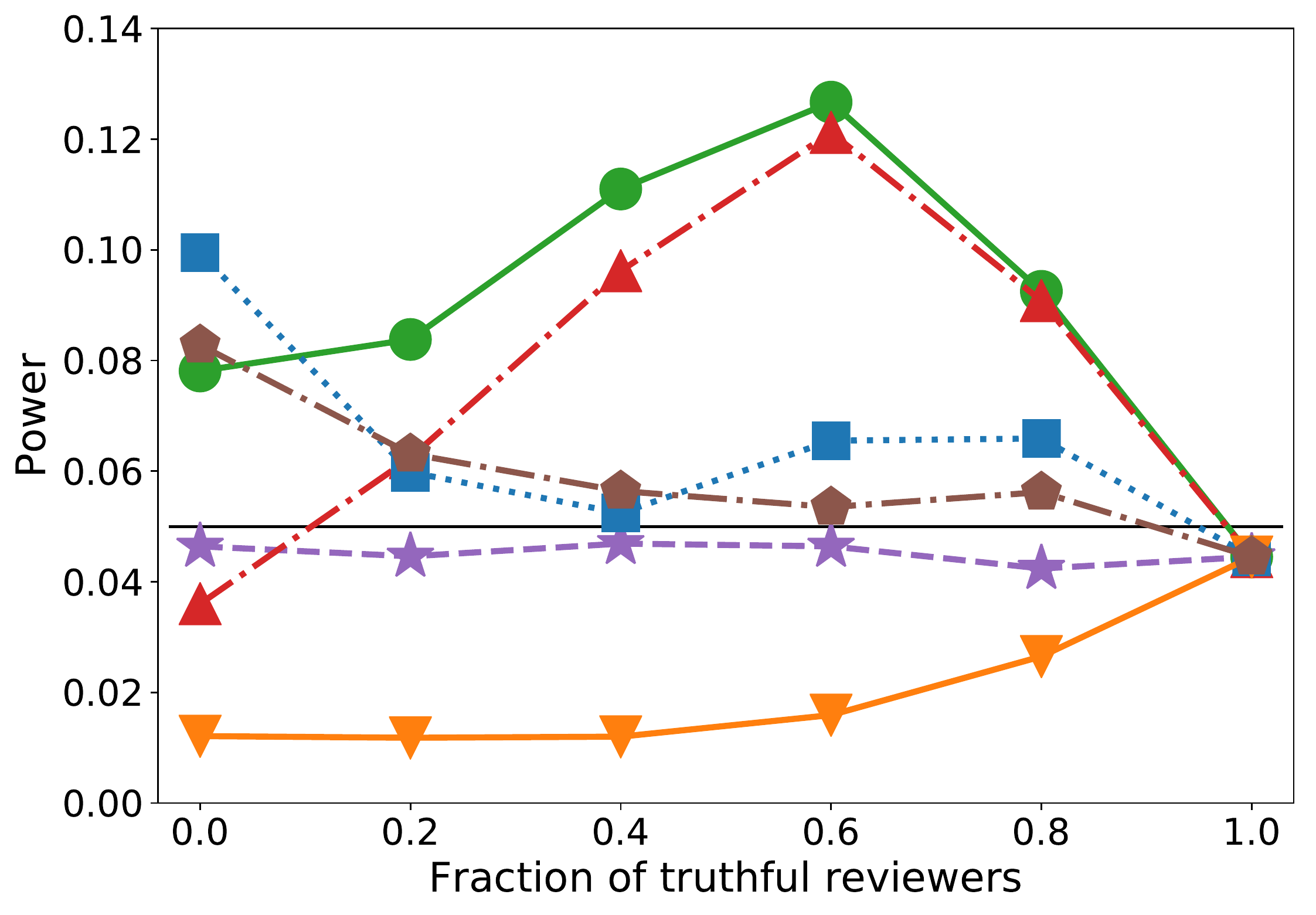}
  \caption{Without supervision}
  \label{fig:power_synthetic_single:t2}
\end{subfigure}
\caption{Detection power of our test against different strategies employed by participants in the experiment. The black horizontal line is a baseline power achieved by a test that rejects the null with probability $\level{=}0.05$ irrespective of the data. Error bars are too small to show.}
\label{fig:power_synthetic_single}
\end{figure}

Figure~\ref{fig:power_synthetic_single} compares the detection power of our test against each strategy used by participants of the experiment. Recall that our test aims at detecting manipulations that improve the final standing of the strategic reviewer. As shown in Figure~\ref{fig:compare_strategies}, the \reverse{} and \second{} strategies do not improve the final standing of the manipulating agent and hence our test cannot detect strategic behaviour when these strategies are employed. 

In contrast, the \heuristic{}, \distance{}, \first{} and \response{} strategies in expectation improve the position of the strategic reviewer when all other players are truthful and hence our test with supervision can detect these manipulations with a non-trivial power, with power being greater for more successful strategies. The behaviour of the test without supervision against these 4 successful strategies involves a complex interplay (depicted in Figure~\ref{fig:power_synthetic_single:t2}) between the fraction of non-strategic agents and the particular strategy employed by strategic players.

%%%%%%%%%%%%%%%%%%%%%%%%%%%%%%%
\subsection{Noisy Supervision}
\label{appendix:moreeval:noise}

The key component of testing with supervision is the set of impartial rankings provided to the test. We now investigate the impact of the noise in the impartial rankings on the power of the test. To this end, we continue with the simulation schema used in the previous section, with the exception that (i) we only consider the \distance{} strategy for the strategic agents and (ii) instead of varying strategies, we vary the level of noise in the impartial rankings. Specifically, we sample impartial rankings from the random utility model using values of the players as quality parameters and adding zero-centered Gaussian noise with standard deviation $\noise$. We then vary parameter $\sigma$ to obtain the power for different noise levels. 

\begin{figure}[ht]
    \centering
    \includegraphics[width=8cm]{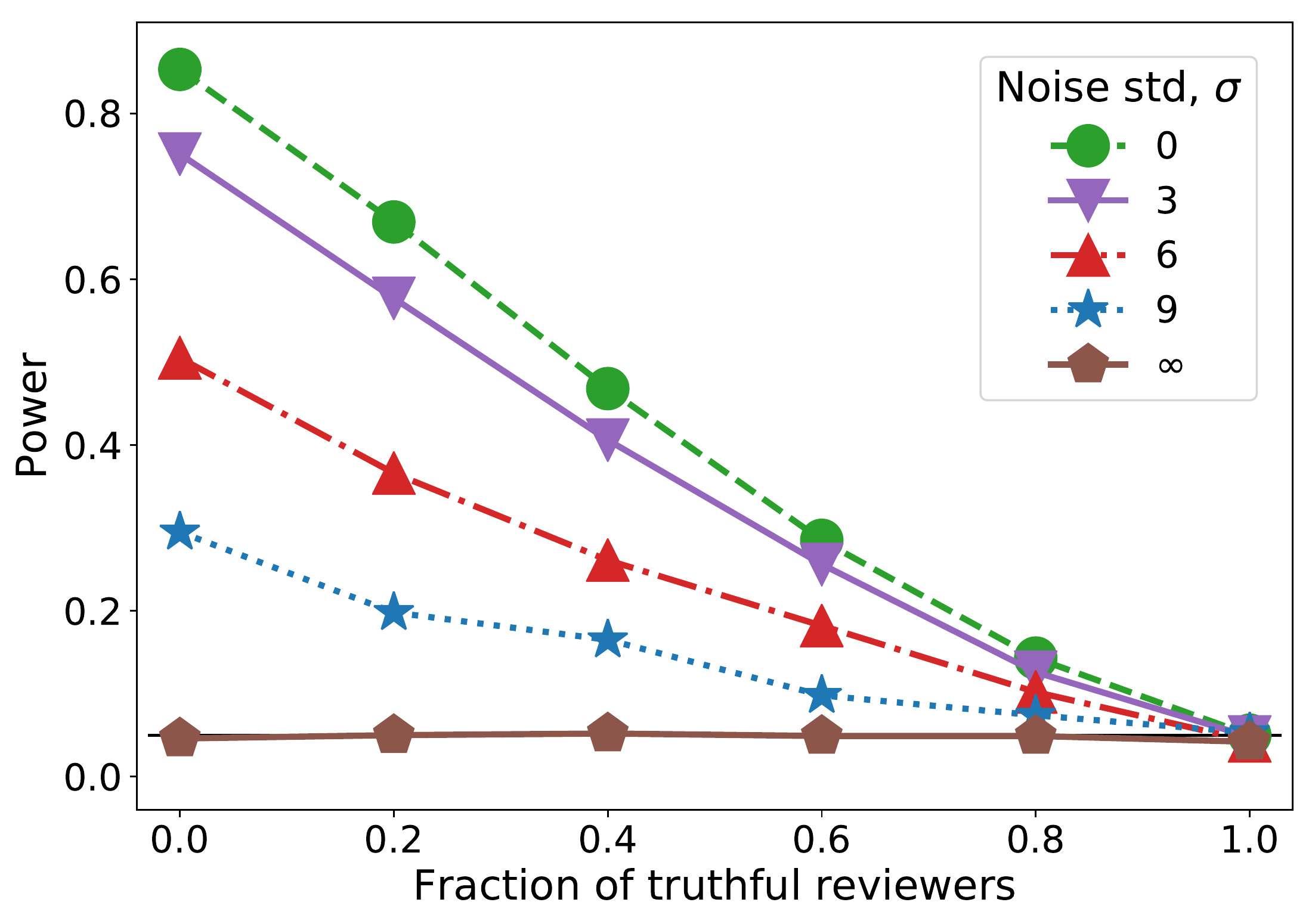}
    \caption{Detection power of our test with supervision for various levels of noise in the impartial rankings. The black horizontal line (hidden behind the line for $\sigma = \infty$) is a baseline power achieved by a test that rejects the null with probability $\level{=}0.05$ irrespective of the data. Error bars are too small to show.}
    \label{fig:power_noise}
\end{figure}

Figure~\ref{fig:power_noise} represents the results of simulations and demonstrates that our test is robust to a significant amount of noise in the impartial rankings. Note that under Gaussian noise with $\sigma = 3$ two players with values differing by 3 points are swapped in the impartial ranking with probability $p \approx 0.24$. Hence, our test with supervision is able to detect manipulations even under significant level of noise. Of course, as the level of noise increases and impartial rankings become random, the power of our test becomes trivial.

%%%%%%%%%%%%%%%%%%%%%%%%%%%%%%%
\subsection{Noise in Reviewers' Evaluations}
\label{appendix:moreeval:noise2}

\begin{figure}[ht]
\centering
\begin{subfigure}[t]{.48\textwidth}
  \centering  
  \includegraphics[height=5.5cm]{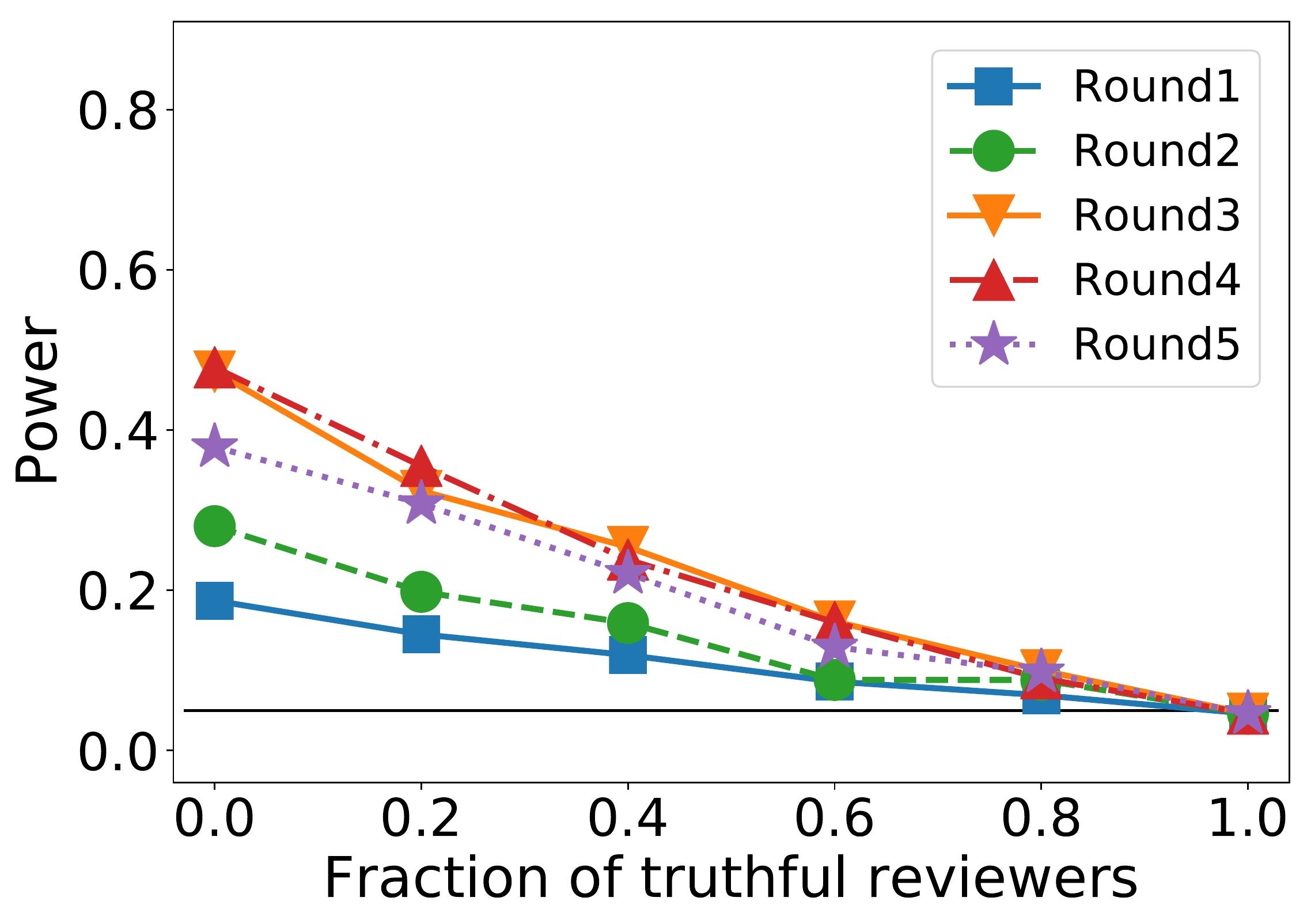}
  \caption{With noiseless supervision}
  \label{fig:power_noise2:t1}
\end{subfigure}%
\begin{subfigure}[t]{.48\textwidth}
  \centering
  \includegraphics[height=5.5cm]{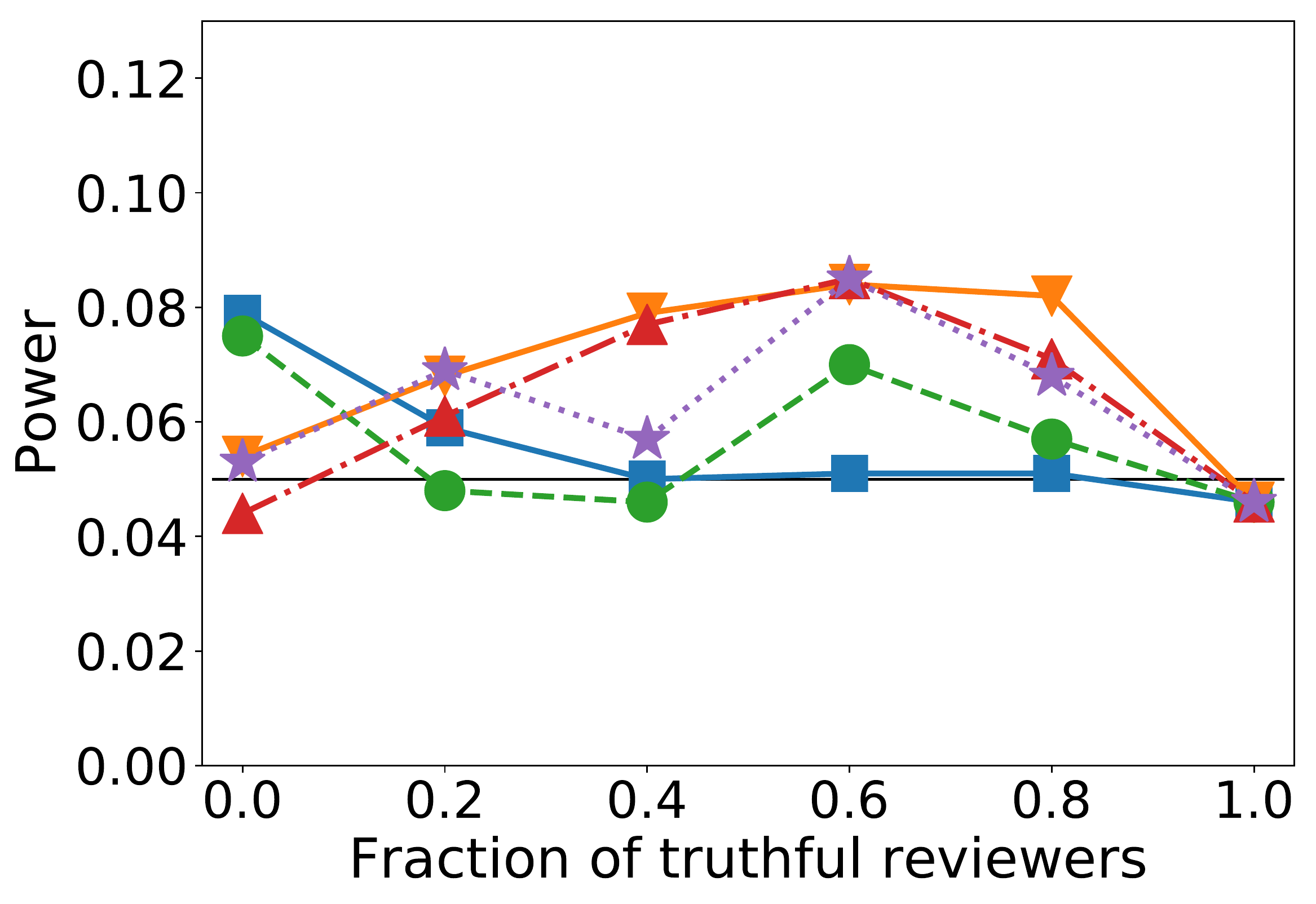}
  \caption{Without supervision}
  \label{fig:power_noise2:t2}
\end{subfigure}
\caption{Detection power of our test when reviewers perceive the ground truth quality of submissions assigned to them with a zero mean Gaussian noise ($\sigma = 3$). The black horizontal line is a baseline power achieved by a test that rejects the null with probability $\level{=}0.05$ irrespective of the data. Error bars are too small to show.}
\label{fig:power_noise2}
\end{figure}

In the experiment we conducted in Section~\ref{sec:experiment} players were communicated exact values of the ground truth quality of submissions, that is, we assumed that reviewers can estimate the quality of submissions without noise. We now study the performance of our test when reviewers are noisy. To this end, we replicate the semi-synthetic setting described in Section~\ref{section:evaluation} with the exception that we add a zero-mean Gaussian noise $(\sigma = 3)$ to scores communicated to artificial agents and compute the detection power of our test under this noisy setup. Figure~\ref{fig:power_noise2} summarizes the results of simulations and shows that out test with supervision continues to have a strong detection power even under significant amount of noise in reviewers' evaluations. Similarly, the test without supervision, while loosing some power as compared to the noiseless case, also manages to maintain a non-trivial power in the noisy case. For additional evaluations of our test when reviewers are noisy we refer the reader to the next section in which some application-specific noise model is considered.

%%%%%%%%%%%%%%%%%%%%%%%%%%%%%%%
\subsection{Violation of Assumption A2}
\label{appendix:moreeval:violation}

The focus of our work is on the peer assessment process and student peer grading is one of the most prominent applications. The literature on peer grading~\citep{piech13moocs, shah13acf} suggests that Assumption A2 that we formulated in Section~\ref{sec:formulation} may be violated in this application: the models proposed in these works (which are used in Coursera and Wharton's peer-grading system) suggest that authors of stronger submissions are also more reliable graders. While our theoretical analysis does not guarantee control over the false alarm probability in this case, we note that in practice our test does not break its respective guarantees under such relationship.

Intuitively and informally, Figure~\ref{fig:compare_strategies} suggests that authors of stronger works benefit from reversing the ranking of submissions assigned to them whereas authors of weaker submissions should play truthfully to maximize the outcome of their submission. In contrast, the aforementioned relationship between the quality of a submission and the grading ability of its author claims the converse: authors of top submissions return rankings that are closer to the ground truth than noisy rankings returned by authors of weaker submissions. Hence, truthful reviewers do not benefit from the difference in noise levels, suggesting that our test may be more conservative, but will not break its false alarm guarantees.

To validate this intuition, we consider the problem parameters given in~\eqref{eqn:simparam} and assume that each reviewer $\revidx \in \revset$ samples the ranking of the works assigned to them from the random utility model with reviewer-specific noise level $\noise_{\revidx}$ and quality parameters determined by the true values of the works. We then simulate the false alarm probability of our test under two setups with different definitions of noise:

\begin{itemize}[itemsep=0pt, leftmargin=*]
    \item \textbf{Setup 1} If reviewer $\revidx$ is the author of one of the top 10 works, they are noiselesee, that is, $\noise_{\revidx} = 0$. In contrast, if reviewer $\revidx$ is the author of one of the bottom 10 submissions, their noise level is non-zero: $\sigma_{\revidx} = \sigma$.
    
    \item \textbf{Setup 2} Each reviewer $\revidx$ samples the ranking from the random utility model with noise level $\sigma \times \nicefrac{k_{\revidx}}{20}$, where $k_{\revidx}$ is the position of the work authored by reviewer $\revidx$ in the underlying ground-truth ordering.
\end{itemize}

\begin{figure}[t]
\centering
\begin{subfigure}[t]{.48\textwidth}
  \centering  
  \includegraphics[height=4.5cm]{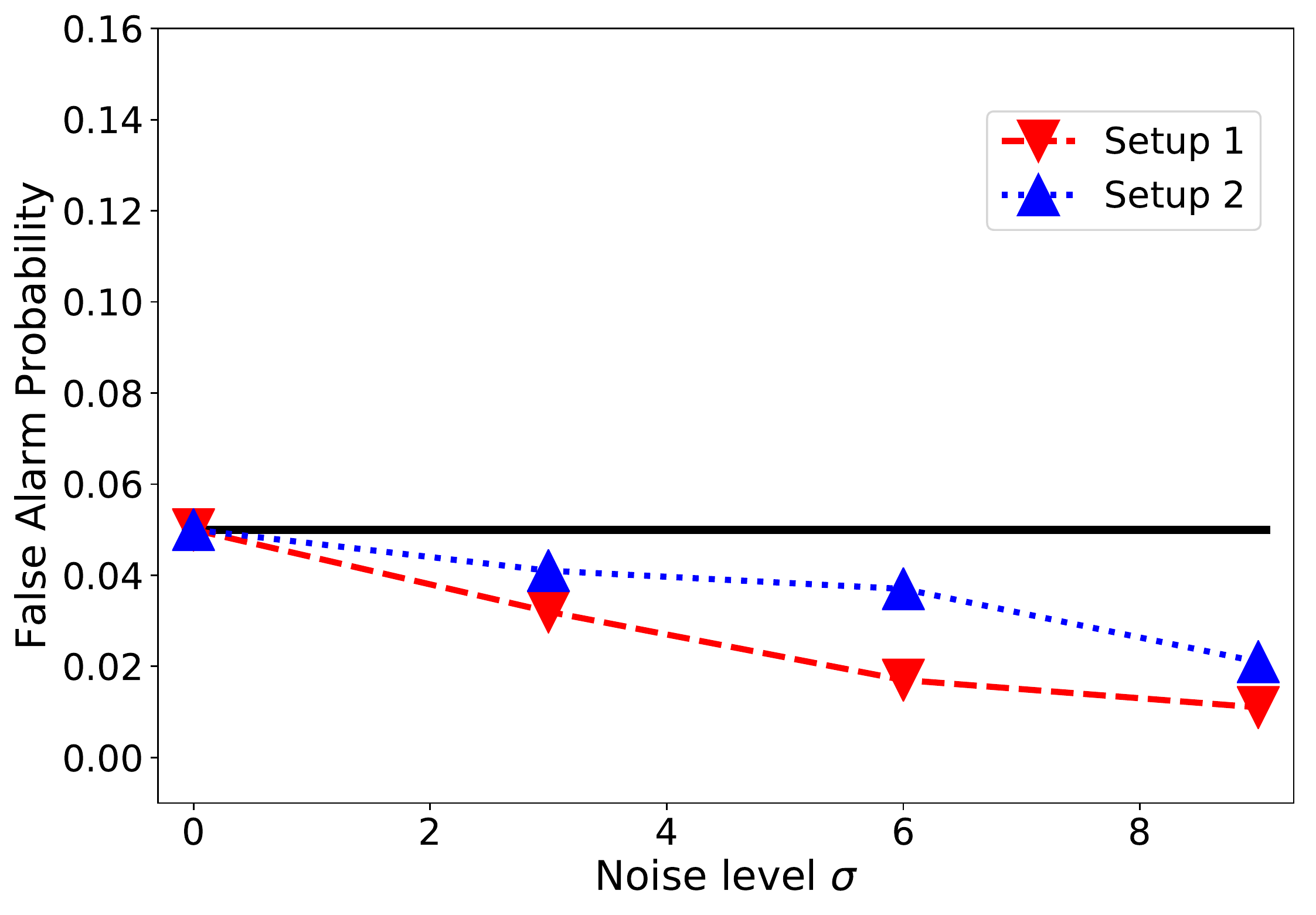}
  \caption{With supervision}
  \label{fig:violation:t1}
\end{subfigure}%
\begin{subfigure}[t]{.48\textwidth}
  \centering
  \includegraphics[height=4.5cm]{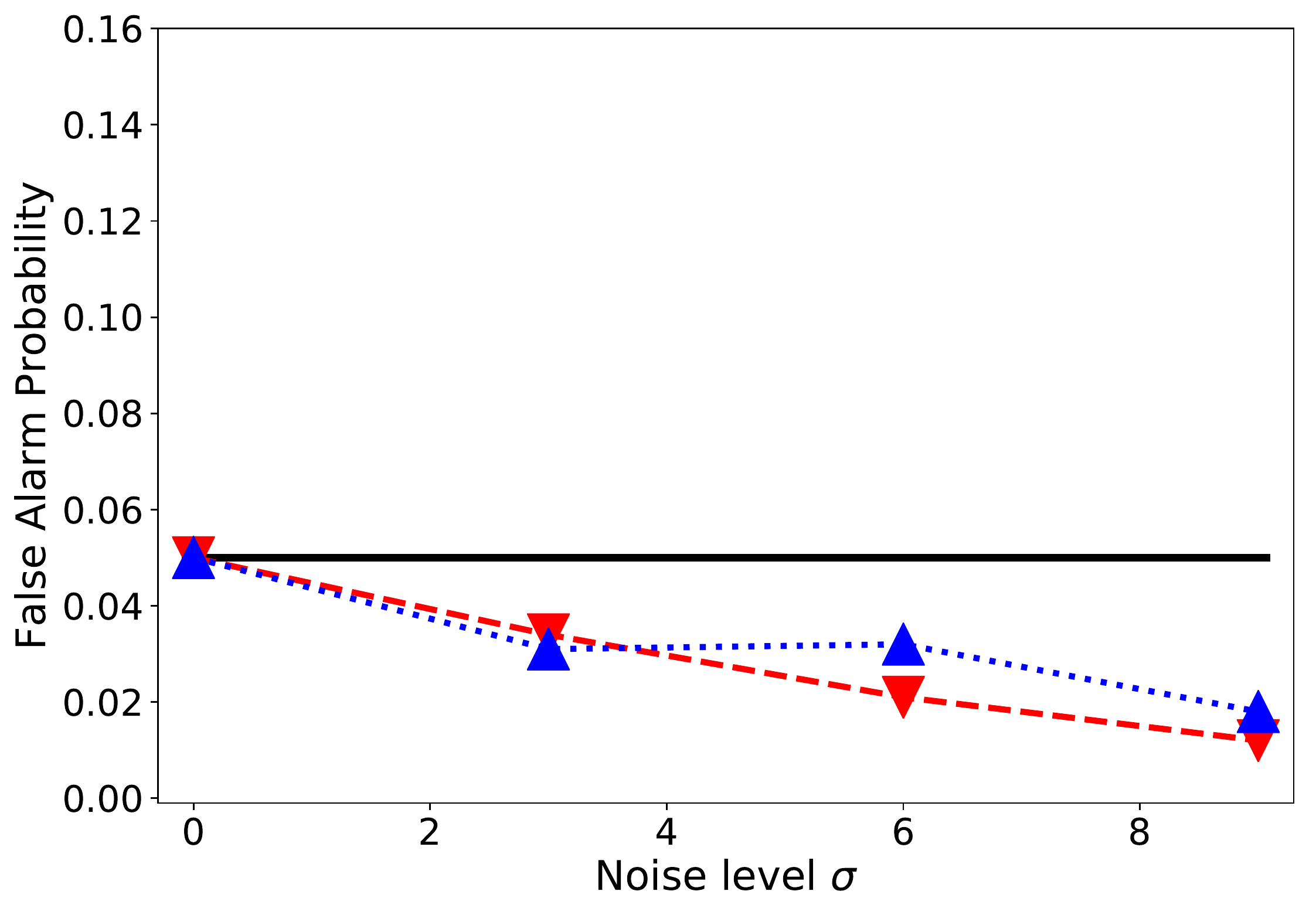}
  \caption{Without supervision}
  \label{fig:violation:t2}
\end{subfigure}
\caption{False alarm probability of our test at the level $\level = 0.05$ when reviewer's noise depends on the quality of their submission. The black horizontal line represents the maximum false alarm probability that can be incurred by a valid test. Error bars are too small to show.}
\label{fig:violation}
\end{figure}

Observe that in both setups data is generated under the null hypothesis of absence of manipulations. In simulations, we vary the noise level $\noise$ and sample impartial rankings for the test with supervision from the random utility model as described in Section~\ref{appendix:moreeval:noise} with noise level $\nicefrac{\noise}{2}$. Figure~\ref{fig:violation} depicts the false alarm probability of our test both with and without supervision and confirms the above intuition: our test indeed controls the false alarm probability when noise in evaluations decreases as the quality of submission authored by the reviewer increases.

\begin{figure}[ht]
\centering
\begin{subfigure}[t]{.48\textwidth}
  \centering  
  \includegraphics[height=4.5cm]{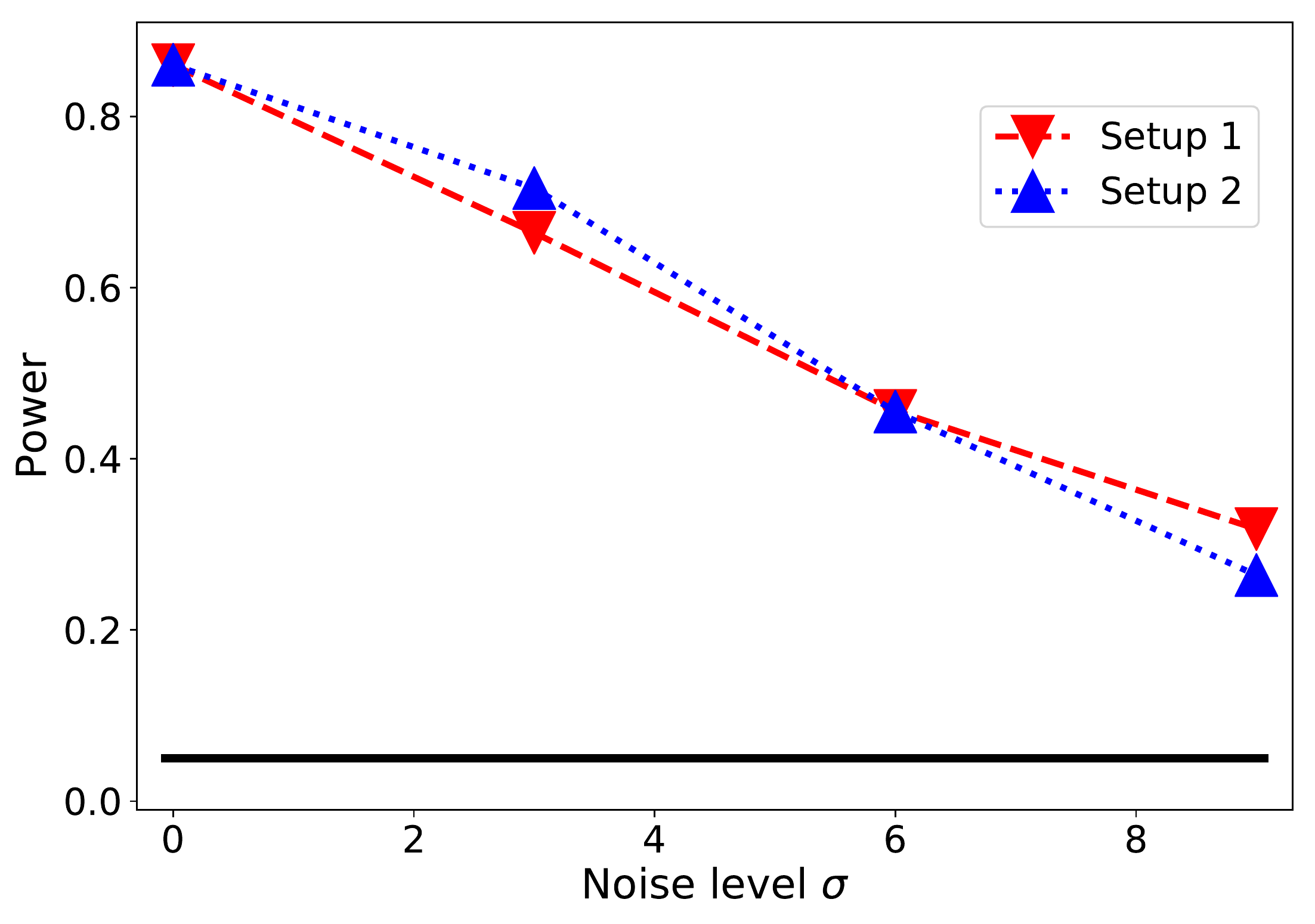}
  \caption{With supervision}
  \label{fig:violation_power:t1}
\end{subfigure}%
\begin{subfigure}[t]{.48\textwidth}
  \centering
  \includegraphics[height=4.5cm]{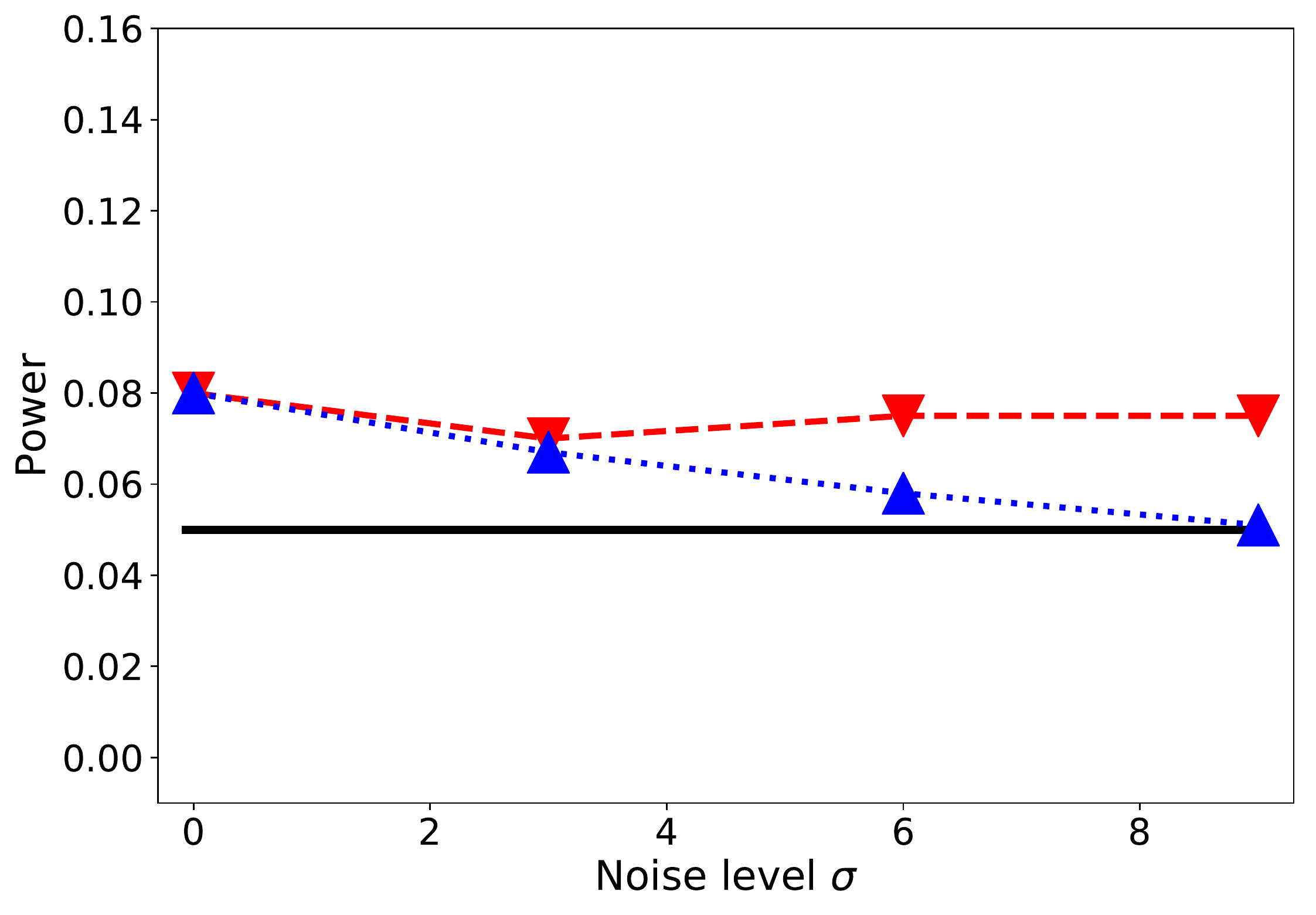}
  \caption{Without supervision}
  \label{fig:violation_power:t2}
\end{subfigure}
\caption{Detection power of our test when reviewer's noise depends on the quality of their submission and all reviewers use the \distance{} strategy. The black horizontal line is a baseline power achieved by a test that rejects the null with probability $\level{=}0.05$ irrespective of the data. Error bars are too small to show.}
\label{fig:violation_power}
\end{figure}

Finally, we note that control over the false alarm probability under violation of the Assumption A2 does not come at the cost of trivial power. Figure~\ref{fig:violation_power} depicts the power of our test under the aforementioned setups when all reviewers manipulate using the \distance{} strategy on top of the noisy values of submissions they sample from the corresponding random utility models and confirms that our tests continue to have non-trivial power in this setup.

%%%%%%%%%%%%%%%%%%%%%%%%%%%%%%%
\subsection{Runtime of the Test} 
\label{appendix:moreeval:pl}

\begin{figure}[ht]
    \centering
    \includegraphics[width=0.5\textwidth]{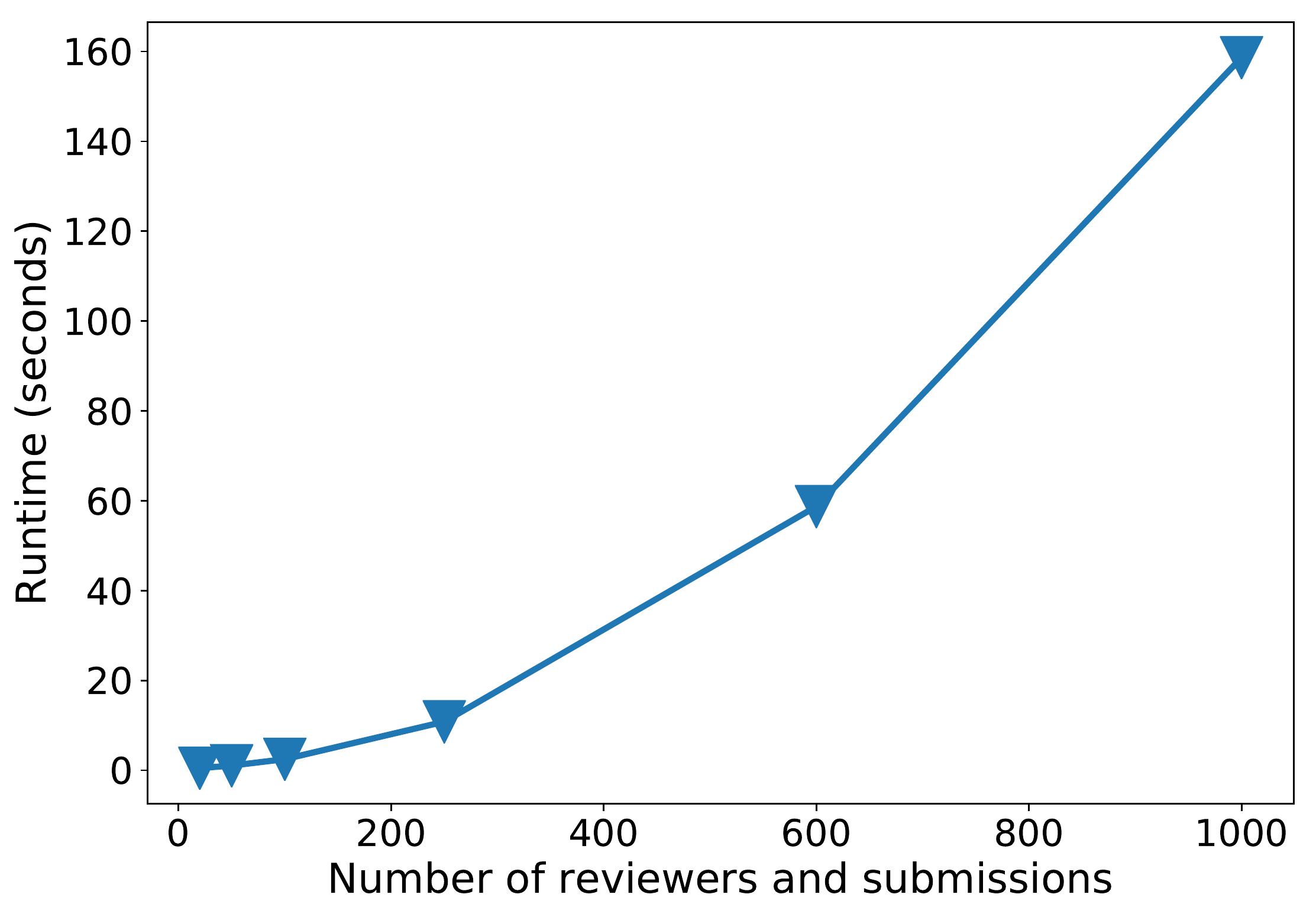}
    \caption{Running time of our test.}
    \label{fig:runtime}
\end{figure}

In this section we continue working with the identity authorship and conflict matrices $(\conflicts = \authorship = \identity)$, considering student peer grading setup in which most of reviewers are conflicted only with their own work. Setting $\papload = \revload = 4$, we estimate the runtime of our test for a wide range of sample sizes $\numpap = \numrev$. Specifically, we use a modification of the test that samples 100 valid authorship matrices in Step~\ref{test1:sample} of Test~\ref{test:withcontrasts} and display the running time in Figure~\ref{fig:runtime}. We conclude that the running time of naive implementation of our test is feasible even for instances with thousands of reviewers and submissions.

%%%%%%%%%%%%%%%%%%%%%%%%%%%%%%%
\section{Random assignment}
\label{appendix:relaxation}

\newcommand{\topology}{T}

When evaluations are collected in the form of rankings, different structures of the assignment graph have different properties that may impact the quality of the final ordering~\citep{shah16topology}. Therefore, one may want to choose a structure of the assignment graph instead of sampling it uniformly at random and we now show how to achieve any desirable structure without breaking the guarantees of our test.

Recall that $\numrev$ is the number of reviewers and $\numpap$ is the number of submissions. Let $\topology$ be the desired structure of the assignment, that is, $\topology$ is a bipartite graph with $\numrev$ nodes in the left part and $\numpap$ nodes in the right part, such that each node in the left part has degree $\revload$ and each node in the right part has degree $\papload$. Given a set of reviewers $\revset$, a set of works $\papset$ and a conflict matrix $\conflicts$, assignment $\assignment$ can be constructed by allocating reviewers and works to the nodes of the graph $\topology$ uniformly at random, subject to the constraint that the assignment $\assignment$ does not violate the conflict matrix $\conflicts$. Note that the resulting assignment will by construction have the desired structure $\topology$.

The above procedure assumes that conflict matrix $\conflicts$ admits structure $\topology$, that is, there exists an allocation of reviewers to nodes that does not violate $\topology$. In practice, it is always the case as reviewers are typically conflicted with only a handful of works (e.g., students in class are only conflicted with their own homework). 

Observe that conditioned on topology $\topology$, the key component of the proof of Theorem~\ref{thm:type1contrasts} captured by equation~\eqref{eqn:base} holds by construction of the assignment $\assignment$, thereby ensuring that the assignment constructed in the described way does not brake the guarantees of our test. Overall, the requirement of random assignment procedure can be relaxed to the requirement of random assignment that respects a given topology $\topology$, thereby enabling deterministic selection of the assignment structure.

%%%%%%%%%%%%%%%%%%%%%%%%%%%%%%%
\section{Proof of Theorem~\ref{thm:type1contrasts}}
\label{appendix:proof}

\newcommand{\prob}[1]{\mathbb{P}\left[ #1 \right]}
\newcommand{\randconflicts}{\widetilde{\conflicts}}
\newcommand{\randauthorship}{\widetilde{\authorship}}
\newcommand{\reviewer}{\rho}
\newcommand{\proauthorship}{{\authorship^{\ast}}}
\newcommand{\proconflicst}{{\conflicts^{\ast}}}
\newcommand{\specset}{\mathcal{D}}
\newcommand{\leftmatrix}{L}
\newcommand{\rightmatrix}{R}
\newcommand{\papcontent}{\Gamma}

Recall that $\papset$ is a set of works submitted for review and $\revset$ is a set of reviewers. Let us also use $\proauthorship$ and $\proconflicst$ to denote authorship and conflict matrices up to permutations of rows and columns, that is, actual matrices $\authorship$ and $\conflicts$ satisfy:
\begin{align}
\label{eqn:generator}
    \conflicts  &= \leftmatrix \proconflicst \rightmatrix \\ 
    \nonumber \authorship &= \leftmatrix \proauthorship \rightmatrix
\end{align}
where $\leftmatrix$ and $\rightmatrix$ are matrices of row- and column-permutations, respectively.

Let $\Perm_{\numrev}$ and $\Perm_{\numpap}$ be sets of all permutation matrices of $\numrev$ and $\numpap$ items, respectively. Conditioned on $\papset, \revset, \proconflicst$ and $\proauthorship$, we note that Assumptions A2 (exchangeability of reviewers and works) ensures that the actual pair of conflict and authorship matrices $\left( \conflicts, \authorship \right)$ follows a uniform distribution over the multiset
\begin{align*}
    \specset = \left\{ \left(\leftmatrix \proconflicst \rightmatrix, \leftmatrix \proauthorship \rightmatrix \right) \Big | (\leftmatrix, \rightmatrix) \in \Perm_{\numrev} \times \Perm_{\numpap} \ \right\}.
\end{align*}

We will now show the statement of the theorem for any tuple $(\papset, \revset, \proconflicst, \proauthorship)$, thus yielding the general result. Let $(\randconflicts, \randauthorship)$ be a random variable following a uniform distribution $\uniform(\specset)$ over the set $\specset$. The proof of the theorem relies on the following fact: if assignment matrix $\assignment$ is selected uniformly at random from the set of all assignments valid for the conflict matrix $\randconflicts$, then for any pairs $(\conflicts_1, \authorship_1) \in \specset$ and $(\conflicts_2, \authorship_2) \in \specset$ that do not violate the assignment $\assignment$, it holds that:
\begin{align}
\label{eqn:base}
    \prob{(\randconflicts, \randauthorship) = (\conflicts_1, \authorship_1) \Big| \assignment} = \prob{(\randconflicts, \randauthorship) = (\conflicts_2, \authorship_2) \Big| \assignment},
\end{align}
where probability is taken over the randomness in the assignment procedure and uniform prior over the pair of conflict and authorship matrices. Indeed, it is not hard to see that:
\begin{align*}
    \prob{(\randconflicts, \randauthorship) = (\conflicts_1, \authorship_1)  \Big| \assignment} &= \frac{\prob{\assignment \Big| (\randconflicts, \randauthorship) = (\conflicts_1, \authorship_1) } \prob{(\randconflicts, \randauthorship) = (\conflicts_1, \authorship_1)}}{\prob{\assignment}} \\ &= \frac{\prob{\assignment \Big| (\randconflicts, \randauthorship) = (\conflicts_2, \authorship_2)} \prob{(\randconflicts, \randauthorship) = (\conflicts_2, \authorship_2)}}{\prob{\assignment}} \\ &= \prob{(\randconflicts, \randauthorship) = (\conflicts_2, \authorship_2)  \Big| \assignment},
\end{align*}
where the second equality follows from the fact that
\begin{align*}
    \prob{\assignment \Big | (\randconflicts, \randauthorship) = (\conflicts_1, \authorship_1)} = \Bigg|\left\{ \assignment' \Big |  \assignment' \text{ is a valid assignment for } \conflicts_1 \right\} \Bigg|^{-1}
\end{align*}
and a simple observation that for any conflict matrix $\conflicts'$ that satisfies~\eqref{eqn:generator} the number of valid assignments is the same by symmetry.

Equation~\eqref{eqn:base} ensures that given the assignment matrix $\assignment$, the randomness of the assignment procedure induces a uniform posterior distribution over the multiset $\conflictset(\assignment)$ of authorship matrices that do not violate the assignment $\assignment$ which we construct in Step~\ref{test1:distr} of the test. Therefore, the actual authorship matrix $\authorship$ is a sample from this distribution: $\authorship \sim \uniform(\conflictset(\assignment))$.

Conditioned on the assignment $\assignment$, under the null hypothesis, for each reviewer $\revidx \in \revset$ the ranking $\ranking_{\revidx}$ is independent of works $\authorship(\revidx)$ and is therefore independent of $\authorship$. Additionally, the optional impartial rankings $\{\impranking_{\revidx}, \revidx \in \revset \}$ are independent of the authorship matrix $\authorship$ by definition. Combining these observations, we deduce that conditioned on the assignment $\assignment$, the test statistic $\stat$ has a uniform distribution over the multiset:
\begin{align*}
    \valset = \left\{ \tmpstat(\tmpauthorship)  | \ \tmpauthorship \in \conflictset(\assignment) \right\},
\end{align*}
where $\tmpstat$ is defined as
\begin{align*}
    \tmpstat(\tmpauthorship) = \sum\limits_{\revidx \in \revset} \sum\limits_{\papidx \in \tmpauthorship(\revidx)} \left( \aggrule_{{\papidx}}(\tmpranking_{1}, \tmpranking_{2}, \ldots, \ranking_i, \ldots, \tmpranking_{\numrev}) - \expectation_{\randranking \sim \uniform_{\revidx}} \left[ \aggrule_{{\papidx}}(\tmpranking_{1}, \tmpranking_{2}, \ldots, \randranking, \ldots, \tmpranking_{\numrev}) \right] \right).
\end{align*}

We finally observe that in Steps~\ref{test1:sample} and \ref{test1:decision} of the test we reconstruct the set $\valset$ and make decision based on the position of the actual value of the test statistic in the ordering of this set. It remains to note that probability of the event that observed value $\tau$ (sampled uniformly at random from the multiset $\valset$) is smaller than the value of $k^{\text{th}}$ order statistic of the multiset $\valset$ is upper bounded by $\frac{k-1}{|\valset|}$. Substituting $k$ with $\left(\left \lfloor {\level \abs{\valset}} \right \rfloor + 1\right)$, we conclude the proof.

\end{document}